\documentclass{olplainarticle}
\RequirePackage{tgtermes}
\RequirePackage{newtxtext}
\RequirePackage{newtxmath}
\RequirePackage{bm}
\RequirePackage{endnotes}

\usepackage{algorithm}
\usepackage{algpseudocode}
\usepackage{tikz}
\usepackage[]{hyperref}
\usepackage{booktabs}
\usepackage{makecell}
\usepackage{multirow}
\usepackage{multicol}
\usepackage[nameinlink]{cleveref}
\usepackage{siunitx}
\usepackage{xspace}
\usepackage{subcaption}
\usetikzlibrary{shapes.geometric,arrows.meta,positioning,calc}

\usepackage{pgfplots}
\pgfplotsset{compat=1.16}

\definecolor{pscolor}{RGB}{0, 155, 0}

\definecolor{mlcolor}{RGB}{219, 48, 122}

\newcommand{\loud}[0]{LOUD\xspace}
\newcommand{\karri}[0]{KaRRi\xspace}
\newcommand{\mtkarri}[0]{Mt-KaRRi\xspace}

\newcommand{\myparagraph}[1]{\paragraph*{#1.}}

\newcommand{\Gveh}[0]{\ensuremath{G}}
\newcommand{\Vveh}[0]{\ensuremath{V}}
\newcommand{\Eveh}[0]{\ensuremath{E}}
\newcommand{\Gpsg}[0]{\ensuremath{G_{\text{psg}}}}
\newcommand{\Vpsg}[0]{\ensuremath{V_{\text{psg}}}}
\newcommand{\Epsg}[0]{\ensuremath{E_{\text{psg}}}}

\newcommand{\dist}[0]{\ensuremath{\delta}}

\newcommand{\orig}[0]{\text{orig}}
\newcommand{\dest}[0]{\text{dest}}
\newcommand{\Pickups}[0]{\ensuremath{P}}
\newcommand{\Dropoffs}[0]{\ensuremath{D}}

\newcommand{\ins}[0]{\ensuremath{\mathcal{I}}}

\newcommand{\veh}[0]{\ensuremath{\nu}}
\newcommand{\capacity}[0]{\ensuremath{\text{cap}}}

\newcommand{\numstops}[1]{\ensuremath{k(#1)}}
\newcommand{\numstopsveh}[0]{\numstops{\veh}}

\newcommand{\laststop}[1]{\ensuremath{s_{\numstops{#1}}}}
\newcommand{\laststopveh}[0]{\laststop{\veh}}

\newcommand{\lwalkmax}[0]{\ensuremath{l_{\text{walk}}^{\text{max}}}}
\newcommand{\walkspeed}[0]{\ensuremath{v_{\text{walk}}}}

\newcommand{\ddown}[0]{\ensuremath{\delta^{\downarrow}}}

\newcommand{\treq}[0]{\ensuremath{t_{\text{req}}}}

\newcommand{\twaitmax}[0]{\ensuremath{t_{\text{wait}}^{\text{max}}}}
\newcommand{\ttripmax}[0]{\ensuremath{t_{\text{trip}}^{\text{max}}}}
\newcommand{\tservmin}[0]{\ensuremath{t_{\text{serv}}^{\text{min}}}}
\newcommand{\tservmax}[0]{\ensuremath{t_{\text{serv}}^{\text{max}}}}

\newcommand{\tarrmin}[0]{\ensuremath{t_{\text{arr}}^{\text{min}}}}

\newcommand{\tbatch}[0]{\ensuremath{t_{\text{batch}}}}

\newcommand{\tdetour}[0]{\ensuremath{t_{\text{detour}}}}

\newcommand{\ttrip}[0]{\ensuremath{t_{\text{trip}}}}

\newcommand{\ttripplus}[0]{\ensuremath{t_{\text{trip}}^{+}}}

\newcommand{\twalk}[0]{\ensuremath{t_{\text{walk}}}}

\newcommand{\detourweight}[0]{\ensuremath{w_\text{detour}}}

\newcommand{\walkweight}[0]{\ensuremath{w_\text{walk}}}

\newcommand{\s}[0]{\si{\second}}
\newcommand{\ms}[0]{\si{\milli\second}}
\newcommand{\mus}[0]{\si{\micro\second}}

\newcommand{\ST}[0]{\texttt{ST}\xspace}
\newcommand{\KA}[0]{\texttt{KA}\xspace}
\newcommand{\LAzero}[0]{\texttt{LA-0.1\%}\xspace}
\newcommand{\LAone}[0]{\texttt{LA-1\%}\xspace}
\newcommand{\LAfive}[0]{\texttt{LA-5\%}\xspace}
\newcommand{\LAten}[0]{\texttt{LA-10\%}\xspace}
\newcommand{\LAtwentyfive}[0]{\texttt{LA-25\%}\xspace}
\newcommand{\LAfifty}[0]{\texttt{LA-50\%}\xspace}
\newcommand{\LAhundred}[0]{\texttt{LA-100\%}\xspace}

\DeclareRobustCommand\activityloopnode{\tikz[baseline=-1mm]{\node[minimum size=5pt,draw,circle,inner sep=1pt,outer sep=0pt]{$\ast$};}}

\usepackage{natbib}
 \bibpunct[, ]{(}{)}{,}{a}{}{,}%

\newcommand{\FIGURE}[4]{%
\centering
#1
\caption[#2]{#3}
#4
}

\newcommand{\TABLE}[4]{%
\centering
  \caption[#1]{#3}
  #4
  #2
}

\begin{document}

\title{Advancing Dynamic Ride-Pooling Simulation -- A Highly Scalable Dispatcher}

\author{%
\fontsize{11}{11}\selectfont Moritz Laupichler, Robin Andre, Kim Kandler, Peter Sanders, Peter Vortisch

\smallskip

\fontsize{10}{11}\selectfont \{moritz.laupichler, robin.andre, kim.kandler, sanders, peter.vortisch\}@kit.edu

\smallskip

Karlsruhe Institute of Technology, Karlsruhe, Germany
}

\begin{abstract}
In ride-pooling, a fleet of vehicles is dynamically dispatched to bring travelers from A to B, trying to pool riders with similar itineraries to improve the use of resources compared to taxis or private cars.
Ride-pooling is considered a core building block of future transport systems with autonomous vehicles.

In this paper, we introduce \mtkarri, a novel dispatcher for dynamic ride-pooling that leverages state-of-the-art shortest-path algorithms to process millions of travelers per hour.
We add a simple mode choice model and use realistic travel demand in three different urban areas for extensive experiments.
We find that our dispatcher scales well with a response time per request of around $1\ms$ even for our largest instances.
We show how this scalability can be used to conduct ride-pooling studies at unprecedented scale.
For instance, we determine how the quality of rides and usage of vehicle resources develop for tens of thousands of vehicles and millions of travelers.

We envision \mtkarri as a tool for future ride-pooling simulation studies at scale.
\end{abstract}


\funding{This paper was created in the ``Country 2 City - Bridge'' project of the ``German Center for Future Mobility'', which is funded by the German Federal Ministry of Transport.}

\maketitle


\section{Introduction}
\label{sec:introduction}
In current transportation systems, travelers usually have a choice between walking, cycling, using a private car, and traditional public transit with fixed routes and schedules. 
Each of these modes is best suited for a different type of journey.
Walking and cycling are easily accessible and compare favorably for short trips.
Public transit is convenient for longer trips in areas with good network coverage, usually within or between larger cities.
Private cars are the most flexible and allow door-to-door journeys for arbitrary origin and destination locations.
However, private cars are resource-inefficient due to the large amount of unused seating capacity.
For example, in Germany, an average car trip has only \num{1.5} passengers~\citep{Belz2025} with similar rates in other Western European countries~\citep{Eurostat2018}. 
Thus, cars take up a disproportionally large amount of space in urban environments and contribute to noise pollution and carbon emissions.
Although scheduled public transit is much more efficient in these respects, it is often unable to provide a viable alternative to cars due to a lack of network coverage and convenience.

Recently, there has been an increasing interest in novel modes of transportation that combine the flexibility of cars with the efficiency of public transit by providing on-demand service in shared vehicles without fixed routes or schedules.
\emph{Ride-pooling} describes a service in which a central agency manages a fleet of vehicles and matches travelers to the vehicles.
Each request is assigned to exactly one vehicle that is tasked with picking up the traveler at their origin and bringing them to their destination.
The system can assign multiple travelers to the same vehicle if they have similar or overlapping itineraries.
This makes ride-pooling more efficient than traditional taxis or ride-hailing services such as Uber or Lyft, at the cost of longer journeys caused by detours for other riders.
Due to its flexibility compared to traditional transit, ride-pooling can fill the gaps in transit systems and offer efficient shared rides to travelers who, in the current system, would have no option but to take a car~\citep{ITF2015}. 
In addition, ride-pooling is vital for the effective use of autonomous vehicles, as research shows that the introduction of autonomous vehicles threatens to \emph{increase} the number of vehicles on the road and the total vehicle kilometers driven, unless vehicles are shared between travelers~\citep{Fulton2017, Garus2025}.

Most research on ride-pooling is concerned with the potential socio-economic benefits and improvements of the efficiency and quality of transportation systems.
These advantages have been studied in numerous real-world field tests~\citep{Gilibert2020, Kostorz2021, Yu2017, Weckstroem2018, Jokinen2019, Chichung2008, Gargiulo2015, Zhu2021} and simulation studies~\citep{Bischoff2017, Kuehnel2023, Agatz2011, Fagnant2014, Wilkes2021, Zwick2022}.
Oftentimes, the scale of these studies is limited to small areas of operation and small ride-pooling vehicle fleets due to the limited scalability of the dispatcher, i.e., the algorithm that computes the assignments of travelers to vehicles.
Additionally, existing work may not fully appreciate the dynamic nature of the problem by assuming known fixed travel times in a road network, or assuming a-priori knowledge of all future traveler requests (\emph{static ride-pooling}) instead of requests being revealed when they are issued (\emph{dynamic ride-pooling}).
Most simulation studies do not consider ride-pooling in the context of a larger transportation system, and ignore important refinements like optimized meeting points between riders and vehicles.

\myparagraph{Contributions}
In this work, we present the first dispatching algorithm for dynamic ride-pooling that scales to millions of requests per hour and millions of independent ride-pooling vehicles, which represents problem instances that are at least an order of magnitude larger than previous work.
We demonstrate how our dispatcher can serve as the basis of detailed studies of ride-pooling at a massive scale.
We summarize our contributions as follows:
\begin{itemize}
    \item We parallelize the dynamic ride-pooling dispatcher \karri to further improve its scalability.
    \item We combine our dispatcher with a mode choice model to evaluate ride-pooling as part of a transportation system with multiple other modes of transportation (car, walking, public transit).
    \item We create realistic problem instances with up to tens of millions of traveler journeys. 
    We include rider-specific maximum walking distances as well as a categorization between urban and rural origin and destination locations for a detailed analysis of ride-pooling in different contexts.
    \item In a simulation study, we analyze the effects of the vehicle fleet size, customer base, use of meeting points, and the distinction between urban and rural journeys on ride-pooling.
    \item We discuss how our new dispatcher facilitates the study of ride-pooling at an unprecedented scale and give directions on how to use it in conjunction with more sophisticated traffic simulations.
\end{itemize}

\myparagraph{Paper Outline}
The rest of this work is structured as follows.
In~\Cref{sec:related_work,sec:preliminaries}, we describe related work in the field of ride-pooling, introduce our ride-pooling model, and describe the \karri dispatcher on which this work is based.
In~\Cref{sec:simulation}, we describe how we extend \karri with a mode choice model and use multi-threading for improved performance.
In~\Cref{sec:input_data_and_methodology}, we detail the problem instances and quality metrics used in our simulation study.
In~\Cref{sec:experimental_results}, we evaluate the performance and solution quality of our dispatcher in extensive experiments.
Finally, in~\Cref{sec:conclusion}, we summarize the merits of our new dispatcher for future research on ride-pooling and discuss further optimizations and extensions of the dispatching algorithm.

\section{Related Work}
\label{sec:related_work}
Dynamic ride-pooling (also called \emph{taxi (ride-)sharing}, \emph{shared taxis}) has been studied as a way to improve urban mobility by allowing multiple passengers to share a vehicle along overlapping routes. Early work focused on real-time matching of requests with available vehicles while minimizing detours, waiting times, and operational costs. \cite{ota_scalable_2015} proposed a foundational online dynamic ride-pooling framework, modeling vehicles as dynamic agents with heterogeneous constraints, aiming to minimize incremental travel costs.
\cite{shuo_ma_t-share_2013, ma_real-time_2015} extended these early models to city-scale operational systems. The 2013 system performs insertion feasibility checks to efficiently integrate new requests, while the 2015 study leverages mobile-cloud architectures and spatio-temporal indexing for large-scale, real-time operations. Both assume door-to-door service and emphasize minimizing detours while maintaining service quality.  
Optimization-based approaches have been introduced to address computational complexity in real-time environments. \cite{cao_sharek_2015} developed the SHAREK framework, which reduces the search space through a multi-step filtering method that selects the most relevant ride options. \cite{shah_neural_2020,dehghan_enhanced_2025} utilize approximate dynamic programming to generate feasible assignments, balancing system-wide objectives with operational constraints. 
Agent-based and simulation studies further support system evaluation under realistic demand and network conditions. \cite{zwick_ride-pooling_2021,zwick_ride-pooling_2022} demonstrate the use of large-scale simulations and machine-learning-based demand prediction to assess performance across cities of varying density. \cite{narayan_scalability_2022} apply agent-based modeling to over 168,000 agents, highlighting system-level efficiency and constraints. \cite{danassis_putting_2022} provides a survey of taxi-data-based models, which presents the current research.

Scalability is a key challenge in dynamic ride-pooling, particularly for dense urban areas with high trip volumes. Early methods, such as those by \cite{ota_scalable_2015}, demonstrate linear growth in computational requirements with the number of trips and vehicles, using parallel processing to maintain city-scale feasibility. \cite{shuo_ma_t-share_2013, ma_real-time_2015} employ lazy shortest-path searches and mobile-cloud architectures to support tens of thousands of vehicles in real-time, while \cite{cao_sharek_2015} achieves millisecond-level response times through multi-stage pruning. \cite{shah_neural_2020} improves fleet efficiency by grouping requests into time batches and assigning vehicles using a linear programming approach. \cite{chen_data_2017} achieve scalability through strict temporal batching (5-minute windows) and geometric pruning of pooling candidates, validated with data from three US cities. 
Aggregation and approximation methods further reduce problem dimensionality. \cite{paparella_time-invariant_2025} apply network-flow aggregation over users and OD pairs, achieving polynomial-time solvability. \cite{liu_bus_2019} use capacitated clustering and location allocation to support city-scale deployment, while \cite{bilali_analytical_2020} model predefined access points to reduce simulation complexity. \cite{soza-parra_shareability_2024} demonstrate that controlling the number of attraction centers (points of interest) and demand structure affects shareability and scalability. 
Empirical evaluations show the behavior of large-scale systems. \cite{zwick_ride-pooling_2021} find that efficiency increases logarithmically with trip density, while \cite{zwick_ride-pooling_2022} confirm transferability of machine-learning-based demand prediction across cities. Furthermore, the research of \cite{shulika_spatiotemporal_2024} presents non-linear scaling with rising demand, which stabilizes after reaching a critical mass.

Maintaining service quality in low-density or peripheral areas is challenging. Early approaches, such as \cite{ota_scalable_2015}, reject requests when detour or waiting constraints cannot be met. \cite{paparella_time-invariant_2025} enforces maximum waiting and delay thresholds, which are harder to satisfy in sparsely populated areas. \cite{dehghan_enhanced_2025} and \cite{cao_sharek_2015} impose strict waiting and price constraints to ensure user satisfaction, while \cite{yu_integrated_2020} report longer waiting times and higher rates of unmet demand in peripheral zones. 
Empirical studies confirm that efficiency and service degrade outside dense urban cores. \cite{zwick_ride-pooling_2021, zwick_ride-pooling_2022} show lower pooling efficiency and higher detours in small towns. \cite{engelhardt_quantifying_2019} find that pooling benefits require a minimum adoption rate, with low-density areas dominated by empty mileage. \cite{shulika_spatiotemporal_2024} highlight increased passenger time penalties in the periphery. 
Structural factors contribute to these patterns. \cite{narayan_scalability_2022} observe that peripheral trips are often rejected due to operational limits, resulting in higher empty-drive ratios. \cite{zhu_potential_2022} show that the effectiveness of pooling depends strongly on demand density and batch size, and \cite{soza-parra_shareability_2024} note that increasing the number of attraction centers (points of interest) reduces peripheral shareability due to fewer attractive shared rides and detours.

Most dynamic ride-pooling models assume door-to-door service without designated meeting points. This includes \cite{ota_scalable_2015, shuo_ma_t-share_2013, ma_real-time_2015, cao_sharek_2015, shah_neural_2020, chen_data_2017, yu_integrated_2020, narayan_scalability_2022, zhu_potential_2022, shulika_spatiotemporal_2024}. While effective in dense areas, door-to-door service can limit scalability and reduce feasibility in peripheral zones.
Several studies explicitly model meeting points to reduce operational complexity. \cite{lorente_intermodal_2022} incorporate public transport stops as fixed access points, improving feasibility and robustness in low-coverage areas. \cite{zwick_analysis_2020} apply stop-based boarding with walking distance constraints, and \cite{bilali_analytical_2020} use predefined centroids as boarding locations with waiting-time constraints. \cite{liu_bus_2019} optimize pickup and drop-off locations to further improve system performance.
Meeting points offer observed benefits in efficiency and feasibility. \cite{ota_scalable_2015} identify bottlenecks where meeting points could improve scalability. \cite{lorente_intermodal_2022} and \cite{zwick_analysis_2020} show reduced computational complexity and higher operational feasibility. \cite{bilali_analytical_2020} demonstrate that boarding and waiting constraints at fixed points increase shareability, and \cite{zhu_potential_2022} find that structural access points can help achieve socially optimal pooling under varying demand conditions, such as time-of-day and policy scenarios.

In summary, dynamic ride-pooling systems have matured to handle real-time requests at metropolitan scale, using a combination of algorithmic, approximation, and simulation-based approaches. Scalability is largely achieved in dense urban cores, but service quality degrades in peripheral areas. Most models rely on door-to-door service, while studies incorporating meeting points show improved feasibility, reduced detours, and enhanced performance in low-density zones. This review highlights the state of the art and identifies the challenges of scaling dynamic ride-pooling while maintaining service quality, particularly in peripheral regions.

\section{Preliminaries}
\label{sec:preliminaries}
In this section, we describe how we model a ride-pooling system and the dispatching process.
We outline shortest-path algorithms that are used in the scalable ride-pooling dispatchers that serve as the basis of this work.
We give an overview of the contributions of these dispatchers and how they combine engineered algorithms for road routing with problem-specific speedup techniques.

\subsection{Ride-Pooling Model}
\label{subsec:ride_pooling_model}
In this section, we define the model that underlies our ride-pooling dispatcher.
For this, we summarize the definitions given by~\cite{Laupichler2024} in less formal terms.

\myparagraph{Road Network}
We model a road network as a directed graph $G=(V,E)$ where edges $e \in E$ represent road segments and vertices $v \in V$ represent intersections between all incident edges.
We associate a fixed travel time $\ell(e) = \ell(u,v)$ with every edge $(u,v) \in E$.
Let $\dist(u,v)$ denote the \emph{shortest-path distance} (i.e., travel time) from $u \in V$ to $v \in V$ with respect to $\ell$.

\myparagraph{Vehicle, Stop}
Our dispatcher manages a fleet $F$ of \emph{vehicles} and their routes.
Every vehicle has a fixed seating capacity $\capacity(\veh)$ and a service time interval $[\tservmin(\veh), \tservmax(\veh)]$.
Every vehicle $\veh$ has a current route $R(\veh) = \langle s_0(\veh), s_1(\veh), \dots, s_{\numstopsveh}(\veh) \rangle$ which contains the current or previous stop $s_0(\veh)$ and all scheduled future stops $s_1(\veh), \dots, s_k(\veh)$.
Every stop is associated with an edge in the road network and we make sure that the vehicle always travels the entire length of the edge to facilitate pickups or dropoffs along the road segment.
A vehicle always follows the shortest path between stop $s_i(\veh)$ and the following stop $s_{i+1}(\veh)$.
When a vehicle reaches its next stop $s_1(\veh)$, we remove $s_0(\veh)$ from its route and update the stop indices, so that the former $s_1$ becomes the new $s_0$ and so on. 
Thus, the vehicle is always located somewhere along the shortest path from $s_0(\veh)$ to $s_1(\veh)$.

\myparagraph{Requests}
In dynamic ride-pooling, rider requests are issued in an online manner, i.e. the dispatcher has no knowledge of future requests.
A request $r=(\orig, \dest, \treq)$ consists of an origin location $\orig \in E$, a destination location $\dest \in E$ and a time $\treq$ at which the request is issued.
We dispatch each request to a vehicle as soon as it comes in, i.e., exactly at time $\treq$.
We assume that the earliest possible departure time is also equal to $\treq$.
Thus, we do not allow pre-booking.

\myparagraph{Meeting Points}
Our dispatcher is designed to consider \emph{meeting points}, which are locations in the vicinity of a rider's origin or destination at which the rider can be picked up or dropped off instead of door-to-door service.
Using meeting points may require the rider to walk a short distance, but the detours that vehicles make to service riders become smaller, particularly if the origin and destination are hard to access (e.g., within a neighborhood of streets with low speed limits).

To model walking to and from meeting points, we consider an additional road network $\Gpsg = (\Vpsg, \Epsg)$ consisting of all paths and roads accessible to pedestrians.
We store information about the length in meters of each path segment in $\Epsg$.
Each rider request specifies a maximum walking distance in meters $\lwalkmax(r)$ as well as a walking speed $\walkspeed(r)$.
The set of pickup locations $\Pickups(r)$ and the set of dropoff locations $\Dropoffs(r)$ for request $r$ contain all locations in $\Epsg$ that are at most $\lwalkmax(r)$ away from the origin and destination of $r$, respectively.
Using $\walkspeed(r)$, we get a walking time for each pickup and dropoff location, called \emph{access} and \emph{egress} time, respectively.
The access time also specifies the earliest time by which a pickup can occur at a given pickup location.

Note that this definition of meeting points is more detailed than the definition used by~\cite{Laupichler2024}, since we allow a different walking distance and speed for every rider.

\myparagraph{Insertion}
When a request comes in, our dispatcher evaluates all possible \emph{insertions} of the request into the route of any vehicle.
An insertion is a tuple $\ins=(r,p,d,\veh,i,j)$, describing that vehicle $\veh$ makes a detour between stops $s_i(\veh)$ and $s_{i+1}(\veh)$ to pick up the rider at pickup location $p \in \Pickups(r)$, and makes another detour between stops $s_j(\veh)$ and $s_{j+1}(\veh)$ to drop them off at dropoff location $d \in \Dropoffs(r)$. 
If $i = j$, only one detour is made and the vehicle goes directly from $p$ to $d$ before advancing to $s_{i+1}$.

\myparagraph{Dispatching and Cost Function}
Our dispatcher implements a simple insertion heuristic.
Given a request $r=(\orig, \dest, \treq)$, the dispatcher picks the best insertion according to a cost function and the current state of vehicle routes.
Once the best insertion has been determined, the route state is immediately updated to reflect this insertion before moving on to the next request.

Our insertion cost function is a linear combination of added vehicle operation time, rider trip time, and walking time.
It takes the form
\begin{equation}\label{eq:cost_function}
\begin{aligned}
  c(\ins) &= \ttrip(\ins) + \ttripplus(\ins) + \detourweight \cdot \tdetour(\ins) \\ 
  & + \walkweight \cdot \twalk(\ins) \text{.} 
\end{aligned}
\end{equation}
The trip time $\ttrip(\ins)$ is the total trip time for the new rider according to $\ins$ under the assumption that there will be no future detours of $\veh$ that increase this time.
To account for the increased trip times of other riders that may be caused by the detours for $\ins$, we add the term $\ttripplus(\ins)$, which is the sum of increases in trip time for all affected riders.
The added vehicle operation time $\tdetour(\ins)$ represents the additional time that vehicle $\veh$ spends driving or waiting to facilitate the insertion.
The model parameter $\detourweight$ determines the importance of the detour time in comparison to the rider trip time.
To account for the discomfort of walking, we introduce the cost term $\walkweight \cdot \twalk(\ins)$, which considers the time $\twalk(\ins)$ spent walking to the pickup location or from the dropoff location to the destination. 
The model parameter $\walkweight$ defines the impact of walking discomfort.

We consider a number of constraints that ensure feasibility and provide a guaranteed minimum service quality to riders.
Consider an insertion $\ins = (r, p, d, \veh, i, j)$.
First, $\ins$ must not cause a vehicle's occupancy to exceed its seating capacity at any point along its route.
If the occupancy of vehicle $\veh$ is already equal to its capacity $\capacity(\veh)$ on any leg between $s_i(\veh)$ and $s_j(\veh)$, the insertion $\ins$ is not feasible.
Second, $\ins$ must not cause the vehicle to be active past the end $\tservmax(\veh)$ of its service time interval, i.e., $\tarrmin(\laststopveh) + \tdetour(\ins) \le \tservmax(\veh)$ where $\tarrmin(\laststopveh)$ is the current arrival time at the last scheduled stop. 
Third, a rider should not wait for too long to be picked up.
Assume that a different rider $r'$ of $\veh$ previously accepted a ride with a tentative pickup time $T_p$.
If $\ins$ delays the pickup of $r'$ to later than $T_p + \twaitmax$ (with model parameter $\twaitmax$), then $\ins$ is considered infeasible.
Fourth, the trip times of riders should not become arbitrarily long.
Again, consider an existing rider $r'$ who accepted a ride with a tentative trip time $T_t$.
If $\ins$ increases the trip time of $r'$ to more than $\alpha \cdot T_t + \beta$ (with model parameters $\alpha$ and $\beta$), then $\ins$ is infeasible.
Thus, every rider is guaranteed a latest possible arrival time relative to the trip time they originally accepted.

Note that the form of the cost function given here differs slightly from the original cost function defined by~\cite{Laupichler2024}.
The original cost function uses a weight parameter for the trip time terms instead of the detour term.
Furthermore, the original code uses wait time and trip time constraints based on the request time and direct shortest-path travel time from origin to destination.
At the same time, travelers are offered rides that break these constraints if no other ride is available.
With this, individual riders can restrict vehicles to not allow any detours, making pooling impossible and paralyzing fleets.
Thus, we use constraints relative to the pickup time and trip time of the accepted ride, which ensures that some detours are always possible.

Our cost function can easily be replaced with a function that takes into account other metrics such as monetary cost or emissions. 

\subsection{Shortest-Path Algorithms}
\label{subsec:shortest_path_algorithms}
We summarize the shortest-path (SP) algorithms used by state-of-the-art ride-pooling dispatchers.

\myparagraph{Dijkstra's Algorithm}
Dijkstra's algorithm~\citep{Dijkstra1959} is a SP algorithm for general graphs.
Given a directed graph $G=(V,E)$ with edge weights $\ell(e) \in \mathbb{R}_{\ge 0}$ and a start vertex $s \in V$, the algorithm finds the SP according to $\ell$ from $s$ to every $v \in V$.
The algorithm proceeds by iteratively updating a tentative distance $d[v]$ for every vertex $v \in V$.
Initially, $d[s]=0$ and $d[v] = \infty$ for $v \ne s$.
In every step, the vertex $v$ with the smallest tentative distance is \emph{settled}.
For every neighbor $w \in V$ of $v$, we check whether the distance $d[v] + \ell(v,w)$ to $w$ via $v$ improves the best known distance $d[w]$ to $w$.
If so, $d[w]$ is updated accordingly.
For every settled vertex $v$, $d[v]$ is final and is equal to the SP distance from $s$ to $v$.
To reconstruct paths, we maintain a parent pointer $p[v]$ for every $v \in V$ that points to the vertex $u$ before $v$ on the SP $\langle s, \dots, u, v \rangle$ from $s$ to $v$.

\myparagraph{Contraction Hierarchies (CHs)}
Contraction Hierarchies (CHs)~\citep{Geisberger2012} are the basis of many speedup techniques for SP computations in road networks. 
CHs utilize the inherent hierarchy of roads in a road network by capturing the idea that SPs typically follow increasingly important roads in the beginning and decreasingly important roads in the end.

The CH algorithm conducts two phases.
In the construction phase, the vertices of $G$ are heuristically ordered by importance.
The network is then augmented with shortcut edges such that between any two vertices $u,v \in V$ there is an \emph{up-down SP} from $u$ to $v$, i.e., a path $p = \langle u, \dots, w, \dots, v \rangle$ where (1) vertices from $u$ to $w$ increase in importance and vertices from $w$ to $u$ decrease in importance, and (2) the total distance of $p$ is equal to the SP distance from $u$ to $v$ in $G$.
If the heuristic for vertex importance is chosen well, only a small number of shortcut edges need to be introduced to ensure this property.
In the query phase, this augmented graph is used to find a SP from a source $s \in V$ to a target $t \in V$.
The query runs a Dijkstra search rooted at $s$ that is restricted to upward edges.
When a vertex $u$ is settled, the search only relaxes edges $(u,v)$ for which $v$ is more important than $u$.
Similarly, a reverse Dijkstra search rooted at $t$ is restricted to downward edges.
Whenever the two searches meet, a candidate up-down path from $s$ to $t$ is found, and eventually all possible up-down paths are found.
The construction phase ensures that there is at least one up-down SP, so this approach always finds a SP from $s$ to $t$. 
If the SP contains any shortcut edges, we can transform them to edges in the input graph by storing unpacking information during the construction phase.

\begin{figure}[t]
\FIGURE
{%
    \subcaptionbox{unidirectional Dijkstra\label{fig:search_space_dij}}{\includegraphics[width=0.3\textwidth]{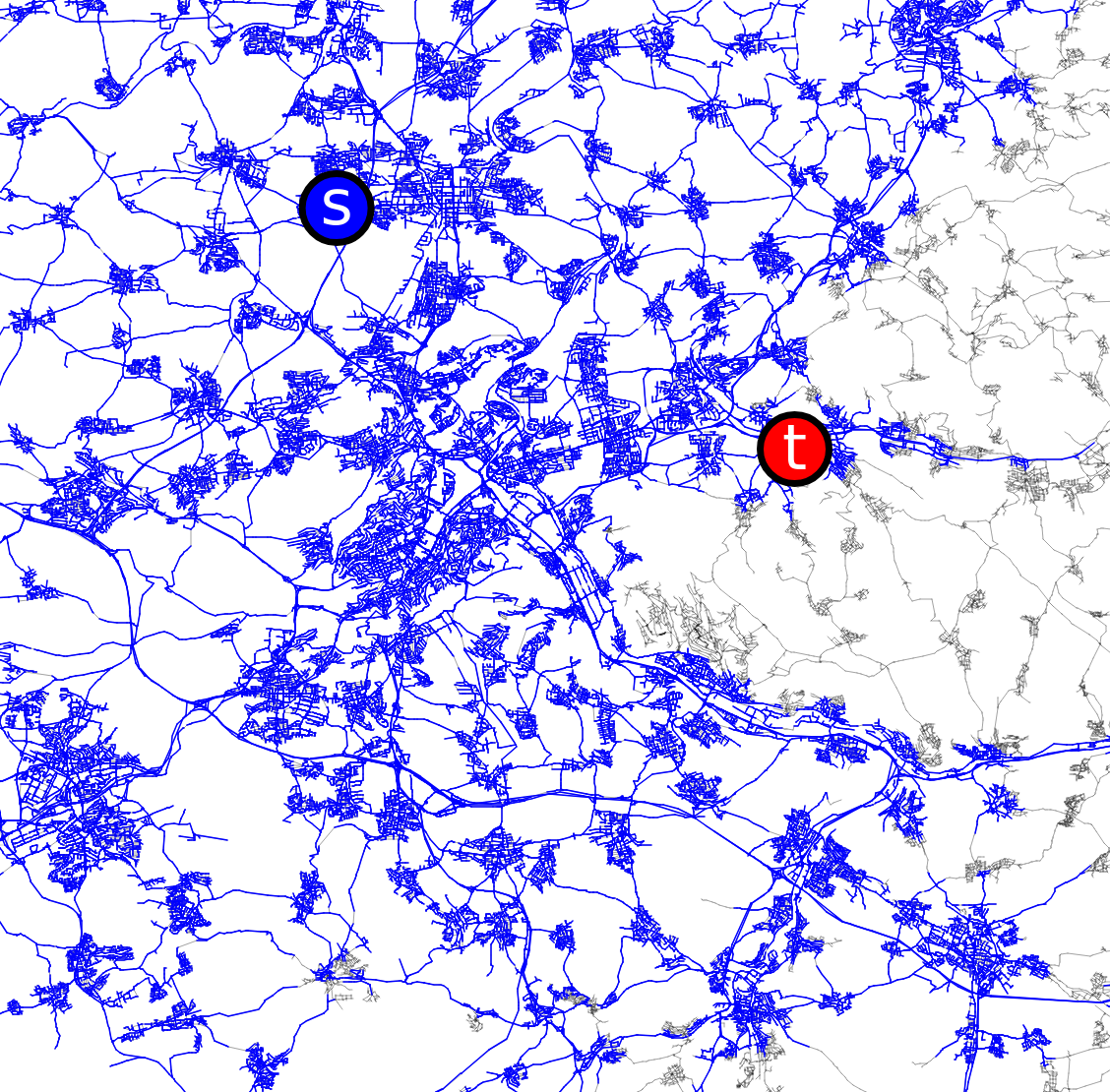}}
    \hfill\subcaptionbox{bidirectional Dijkstra\label{fig:search_space_bidij}}{\includegraphics[width=0.3\textwidth]{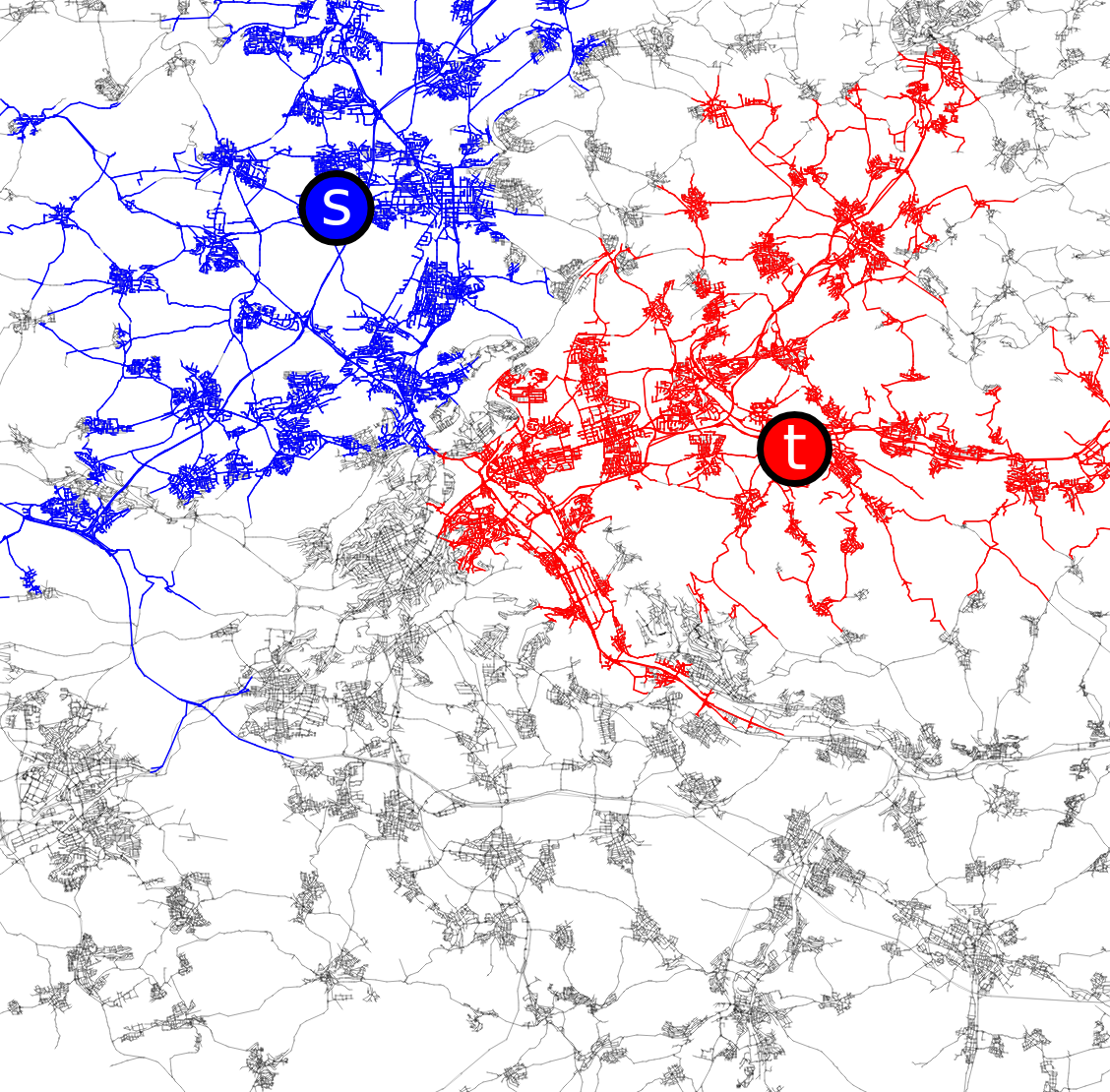}}
    \hfill\subcaptionbox{CH query\label{fig:search_space_ch}}{\includegraphics[width=0.3\textwidth]{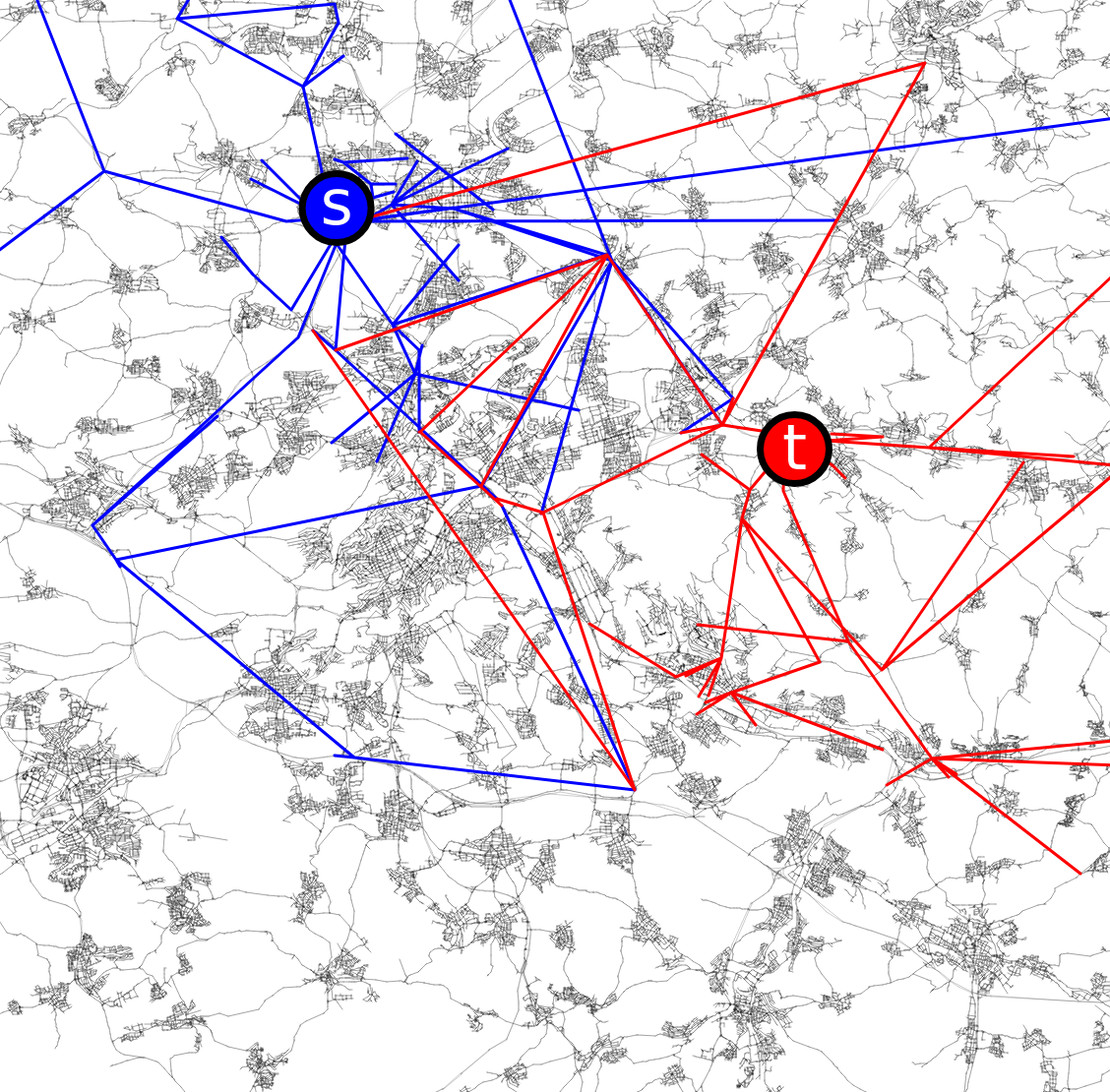}}
}{
    Search spaces of different point-to-point shortest-path algorithms.
}{
    Search spaces for a point-to-point shortest-path query from $s$ to $t$ in the road network of Stuttgart, Germany, using Dijkstra's algorithm (\Cref{fig:search_space_dij}), the bidirectional variant of Dijkstra's algorithm (\Cref{fig:search_space_bidij}), and a CH query (\Cref{fig:search_space_ch}).
    Blue edges mark the search space of the forward query rooted at $s$ and red edges mark the search space of the reverse query rooted at $t$.
    Lines in~\Cref{fig:search_space_ch} include shortcuts which skip over paths of lower importance in the CH.
}{
    \label{fig:search_spaces}
}
\end{figure}
A CH query finds the point-to-point distance from $s \in V$ to $t \in V$ much more efficiently than Dijkstra's algorithm.
Dijkstra's algorithm settles every vertex that is closer to $s$ than $t$.
In comparison, a CH query only needs to settle important vertices and thus search a fraction of the graph.
In~\Cref{fig:search_spaces}, we visualize the search spaces of a SP query from $s$ to $t$ in the road network of Stuttgart, Germany, using Dijkstra's algorithm, the bidirectional variant of Dijkstra's algorithm, and a CH query. 

\myparagraph{One-to-Many Queries using BCHs}
\label{BCH}
The one-to-many SP problem asks to find the SPs from a single source $s \in V$ to every target $t$ in a set of targets $T \subseteq V$.
A technique called \emph{Bucket Contraction Hierarchies (BCHs)}~\citep{Geisberger2012, Knopp2007} utilizes a CH to solve this problem.
Every $v \in V$ has a bucket $B(v)$ which contains entries of the form $(t, \ddown(v, t))$ that specify that there is a path from $v$ to $t \in T$ with length $\ddown(v,t)$ that consists only of downward edges. 

The algorithm proceeds in two phases.
In the selection phase, for each $t \in T$, the algorithm runs a reverse Dijkstra search from $t$ that is restricted to downward edges.
For every vertex $v \in V$ settled, an entry $(v, \ddown(v,t))$ is added to $B(v)$.
Then, in the query phase, the algorithm runs a Dijkstra search from $s$ that is restricted to upward edges.
Whenever this search settles a vertex $v \in V$, each bucket entry $(t, \ddown(v,t)) \in B(v)$ implies an up-down path from $s$ to $t$.
Thus, the up-down SP from $s$ to every $t \in T$ will be found by scanning the bucket entries of every settled vertex.
As long as the set of targets $T$ does not change, the same bucket entries can be used for multiple queries with different sources $s \in V$.
We augment bucket entries with parent pointers to reconstruct full paths.

\subsection{\loud}
\label{subsec:loud}
To determine the best insertion for a new request in dynamic ride-pooling, many shortest-path distances need to be computed (e.g., for vehicle detours).
Many dispatchers simply assume that a table of shortest-path distances for every pair of vertices is known~\citep{Psaraftis1980, Jaw1986, Madsen1995, Horn2002, Hunsaker2002, Haell2012, Ota2017, Lotze2022}.
Although one could pre-compute all pairwise shortest-path distances, this method does not scale to large networks due to memory requirements and the inability to react to changing travel times.
Thus, recent work has been interested in computing the required distances on-the-fly, i.e., when a request comes in, according to up-to-date travel time information in the road network~\citep{Bischoff2017, Horni2016, shuo_ma_t-share_2013, Huang2014, ma_real-time_2015}.

One of these dynamic ride-pooling dispatchers with on-the-fly distance computation is \loud (\underline{Lo}cal B\underline{u}ckets \underline{D}ispatching)~\citep{Buchhold2021}.
\loud determines an insertion for each incoming request according to a cost function and constraints that are similar to the ones used here (see~\Cref{subsec:ride_pooling_model}).
However, \loud does not allow meeting points and does not consider rider trip times in the cost function (other than violations of the trip time constraint).

\myparagraph{Elliptic Pruning}
\loud uses \hyperref[BCH]{BCH} queries to compute the one-to-many distances from a request's origin and destination to the stops in existing vehicle routes. 
More precisely, the algorithm generates bucket entries for every vehicle stop upon adding the stop to the respective vehicle's route.
Then, a BCH query sourced at the origin or destination can use these bucket entries to compute the distances required to determine the cost of potential insertions into the routes.
The bucket entries for a stop can be re-used for incoming requests as long as the stop is part of the vehicle route, i.e. until the vehicle reaches the stop and the stop is removed from the route.

As a main contribution, \loud finds that BCH queries can be pruned, so that they run much faster, using the constraints of existing riders in a vehicle.
\loud identifies a set of vertices (called the \emph{detour ellipse}) to which a detour can be made between any two subsequent stops without violating any constraints.
The authors of \loud show that for stops $s_i$ and $s_{i+1}$, it suffices to generate bucket entries only at a small subset of vertices within the detour ellipse to still find all feasible insertions.
This \emph{elliptic pruning} limits the size of the BCH buckets, which reduces the memory requirements and speeds up bucket scans during queries.
Additionally, only those vehicles for which valid distances are found by the pruned BCH queries can lead to feasible insertions.
Thus, elliptic pruning speeds up both the shortest-path queries and the process of determining the best insertion. 

\subsection{\karri}
\label{subsec:karri}
In this paper, we utilize the dispatching algorithm \karri (\underline{Ka}rlsruhe \underline{R}apid \underline{Ri}de-pooling)~\citep{Laupichler2024}, which addresses a remaining bottleneck of \loud and extends it to allow the use of meeting points.

\myparagraph{Distances to Last Stops in Vehicle Routes}
Elliptic pruning optimizes queries for insertions that insert new stops in between a pair of existing stops of a vehicle route.
However, this is not applicable for the important special case of appending new stops to the end of a vehicle route where there are no more constraints of existing riders.
Thus, \loud uses Dijkstra's algorithm to compute the distances from the last stops in vehicles' routes to the origin and destination of a new request, which makes up a large part of the algorithm's running time.

\karri uses BCHs for this special case, generating bucket entries for the last stop of each route without pruning and querying them for incoming requests.
To avoid the overhead of scanning large numbers of bucket entries, the entries of each bucket are sorted by the associated distance.
Then, based on an upper bound on the cost of the best insertion, the algorithm can stop scanning the entries of a bucket as soon as an entry is reached for which the resulting insertion would be more expensive than this upper bound.
As all remaining entries have a greater distance, they would also incur a higher cost, which means no remaining entry of the bucket can be relevant.
Although the buckets for last stops have many more entries than the pruned buckets, sorting and stopping bucket scans early retains many of the benefits of elliptic pruning.

\myparagraph{Shortest-Path Queries for Meeting Points}
As described in~\Cref{subsec:ride_pooling_model}, meeting points are locations close to the origin or destination location of a vehicle at which the rider may be picked up or dropped off, requiring them to walk from the origin to the pickup location or from the dropoff location to the destination.
For each request, there may be many possible pickup and dropoff locations, and each of these locations leads to a different insertion with different cost.
To compute the costs for all possible insertions, the shortest-path distances between vehicle stops and every pickup and dropoff location must be known.
Thus, the one-to-many shortest-path problems addressed by BCH queries in \loud become many-to-many shortest-path problems.
These can be most easily handled by computing the one-to-many solution for every meeting point.
\karri improves upon this by using the fact that the pickup/dropoff locations are close to each other, which means that large parts of the shortest paths to the vehicle stops are identical for all pickup/dropoff locations.
This allows the algorithm to bundle shortest-path queries, which improves cache access patterns and enables the use of SIMD vector instructions, a special type of instruction on modern processors, which perform computations for multiple searches in parallel.
Additionally, last stop searches can prune the queries for different pickup/dropoff locations by comparing their progress to one another.
This reduces the amount of work needed and acts as a preliminary filter for dominated insertions.

\section{Scalable Ride-Pooling Simulation}
\label{sec:simulation}
In this section, we describe the scalable ride-pooling simulation used in this paper.
First, we give an overview of the workflow of the basic \karri simulation.
Then, we extend \karri with a rider mode choice model, and introduce a parallelized variant of the algorithm.

\subsection{Overview of Ride-Pooling Simulation}
\label{subsec:overview_simulation}
\tikzstyle{decision} = [diamond, draw, aspect=2, inner sep=1pt]
\tikzstyle{action} = [rectangle, rounded corners, draw]
\tikzstyle{data} = [trapezium, draw, trapezium left angle=60, trapezium right angle=-60]
\tikzstyle{merge} = [circle,fill=black,minimum size=2mm,inner sep=0pt]
\tikzstyle{entry} = [circle,draw,minimum size=5mm, inner sep=0pt]
\tikzstyle{exitouter} = [entry]
\tikzstyle{exitinner} = [circle,fill=black,minimum size=3.5mm,inner sep=0pt]
\begin{figure}
\FIGURE
{
    \begin{tikzpicture}[scale=1.5,
    every node/.style={align=center,scale=0.75}, 
    every path/.style={-{>[length=1.5mm,width=2mm]}}]
        \node[decision] (decmore) at (0,2.5) {more events?};
        \node[decision] (dectype) at (0,1.25) {event type?};
        \node[action] (fbi) at (0,0) {find best\\insertion};
        \node[decision] (decf) at (0,-1) {$\exists$ feasible?};
        \node[action] (us) at (0,-2) {update state};
        \node[action] (pv) at (-3,1.25) {process vehicle\\event};
        \node[data] (ss) at (-3,0) {system state};
        \node[merge] (splitus) at (ss |- us) {};
        \node[data] (e) at (-3,2) {events};
        \node[merge, left=0.5cm of decmore] (mergeentry) {}; 
        \node[entry] at ([xshift=-0.8cm,yshift=0.2cm]mergeentry) (entry) {};
        \node[entry] at ([xshift=-0.8cm,yshift=-0.2cm]mergeentry) (entryloop) {\Large $\ast$};
        \node[merge, right=0.5cm of us] (mergedonerec) {}; 
        \node[entry, right=0.5cm of mergedonerec] (donerec) {\Large $\ast$};
        \node[entry] (donepv) at ([xshift=5mm,yshift=-4mm]pv.south) {\Large $\ast$};
        \node[exitouter, right=1cm of decmore] (exit) {};
        \node[exitinner] at (exit) {};
        \draw[] (entry) to [bend left=20] (mergeentry);
        \draw[] (entryloop) to [bend right=20] (mergeentry);
        \draw[] (mergeentry) -- (decmore);
        \draw[] (decmore) -- node[above,pos=0.33] {no} (exit);
        \draw[] (decmore) -- node[right,pos=0.33] {yes} (dectype);
        \draw[] (dectype) -- node[above,pos=0.4] {vehicle} (pv);
        \draw[] (dectype) -- node[right,pos=0.4] {new request} (fbi);
        \draw[] (pv.south -| donepv.north) -- (donepv);
        \draw[] (fbi) -- (decf);
        \draw[] (decf) -- node[right,pos=0.33] {yes} (us);
        \draw[] (decf) -| node[above,pos=0.15] {no} (mergedonerec);
        \draw[] (us) -- (mergedonerec);
        \draw[] (mergedonerec) -- (donerec);
        \draw[dashed] (pv) -- (e);
        \draw[dashed] (pv) -- (ss);
        \draw[dashed] (ss) -- (fbi);
        \draw[dashed] (us) -- (splitus);
        \draw[dashed] (splitus) -- (ss);
        \draw[dashed] (splitus) -| ([xshift=-5mm]ss.west) |- (e);
        \draw[dotted,-{},thick] (e) to [bend right=20] (decmore);
    \end{tikzpicture}
}{
    Workflow of \karri in its basic form.
}{
    Solid arrows indicate control flow.
    Dashed arrows indicate data reads and writes.
    Dotted lines indicate information for decisions.
    The node \activityloopnode{} signifies a loop to the next event.
}{
    \label{fig:workflow_karri}
}
\end{figure}
In its original form~\citep{Laupichler2024}, \karri runs a simple event simulation with vehicle events and request events.
Every event can update the system state, which consists of vehicle routes and the data structures for shortest-path queries.
We illustrate the workflow in~\Cref{fig:workflow_karri}.

For vehicles, there are three types of events: startup, reached stop, and shutdown.
Initially, all vehicles are inactive and the simulation contains a startup event for every vehicle at the beginning of the vehicle's service time.
When a startup event occurs, the vehicle becomes active and a single stop is inserted into the vehicle's route at its initial location.
The algorithm inserts bucket entries for the new stop into the BCH data structures to allow shortest path queries for this stop.
As long as the vehicle has exactly one stop, it idles there until it is assigned a rider, prompting it to move to new stops.
For every vehicle $\nu$ with a non-empty route, the event simulation contains a ``reached stop'' event at the time when $s_1(\nu)$ will be reached. 
Triggering this event means that the vehicle has finished its current leg, which causes the stop $s_0(\nu)$ to become obsolete.
The algorithm removes $s_0(\nu)$ from the route $R(\nu)$ and deletes its bucket entries.
If more than one stop remains in $R(\nu)$, a new ``reached stop'' event is added for the new $s_1(\nu)$ according to the current vehicle schedule.

For every request $r$, the simulation contains a ``new request'' event at the request's time of issue $\treq$.
When this event is reached, the algorithm computes the best insertion based on the current vehicle routes.
It uses the bucket entries for active stops to compute the required shortest paths efficiently (see~\Cref{subsec:loud,subsec:karri}).
If at least one feasible insertion is found, \karri performs the insertion by adding stops to the vehicle's route, either in between existing stops for a detour or appended to the end of the route.
For any new stops, the algorithm generates bucket entries in the BCH shortest-path data structures.
If the affected vehicle $\nu$ was previously idle or the next stop $s_1(\nu)$ has changed, the vehicle immediately starts moving from its current location to the new $s_1(\nu)$ on the shortest path. 
Additionally, a new ``reached stop'' event is added for the new $s_1(\nu)$.

\subsection{Rider Mode Choice Mechanism}
\label{subsec:mode_choice}
For every request, \karri computes the feasible insertion with the smallest cost according to its cost function (see~\Cref{subsec:ride_pooling_model}).
The dispatcher offers a ride to the traveler according to this insertion. 
If a feasible insertion exists at all, the original algorithm assumes that the traveler always accepts the ride offer.
With this assumption, the ride-pooling system serves all travelers, but the vehicles also have to perform every ride, even those with very high overhead.
Thus, vehicle routes and service quality tend to degenerate if demand exceeds the capabilities of the vehicle fleet.

However, it is unrealistic for travelers to always accept the ride offer and use ride-pooling, especially when ride offers get worse due to high utilization of the ride-pooling system. 
Therefore, in this paper, we extend \karri with a simple mode choice mechanism that models a traveler's choice between walking, taking their own car, using traditional line-based public transit, and accepting the ride-pooling offer. 
This mode choice mechanism simply replaces the feasibility check in the workflow (cf. ~\Cref{fig:workflow_karri}) and the system state is updated only if the rider chooses ride-pooling.

\myparagraph{Choice Model}
A discrete choice model is an algorithm to select an element $a$ from a set of alternatives $A$. 
To model the decision process for each request, we assign a utility score $U_a \in \mathbb{R}$ to each mode choice, providing a quantitative representation of the factors driving the decision. 
In this paper, we model the utility of each mode $a$ as a linear combination of a mode-specific constant and a travel time factor $U_a =\alpha_a + t \cdot\beta_a + w \cdot \beta^w_a + acc \cdot \beta^{acc}_a$ where $\alpha_a$ captures the inherent preference, $\beta_a$ the impact of travel time $t$ on the choice attractiveness, $\beta^w_a$ the impact of waiting time $w$ and $\beta^{acc}_a$ the impact of access and egress time to the mode (e.g. walking to a station).
We use a logit model as in \cite{Train2003} where the selection probability of an alternative is calculated by $P{_{a}} =\frac{e^{U_{a}}}{\sum_{i \in A} e^{U_{i}}}$. 
The parameters for the utility functions have been calculated from the German ``Mobilität in Deutschland'' survey from \cite{nobis2018MiD}, so that the parameters best match the decisions observed in the travel diaries in the survey. 
We select a mode by calculating the choice probabilities and sampling from the cumulative distribution using a random number.
We also take into account that a traveler may not have access to every mode.
For example, a traveler may not own a private car or may not have viable public transportation nearby.
In this case, we set the utility of the unavailable mode to $U_a = - \infty$, which leads to a mode probability of $P_a=0$.
This model uses the access, egress, and travel times produced by the routing algorithms at request time. 
In particular, any situational effects that arise after the model selection, such as unplanned ride-pooling detours, traffic congestion or public transport delays are not taken into account.

\myparagraph{Application of Discrete Choice Model}
The discrete choice model requires as input the travel time, waiting time and access and egress time from the origin to the destination of the traveler for every mode of transport considered.
For ride-pooling, we can derive these values from the shortest paths in the vehicle and pedestrian network associated with the best insertion.
We obtain the travel time for a private car and for walking by running a point-to-point CH query (cf.~\Cref{subsec:shortest_path_algorithms}) from the origin to the destination in the vehicle and pedestrian network, respectively. 
For line-based public transit, we run a query using the public transit routing algorithm ULTRA~\citep{Baum2023} on a public transit network of the area considered.
We use the bounded multi-criteria version of the algorithm to obtain an approximately-optimal Pareto set of public transit journeys according to the criteria travel time, number of transfers, and transfer walking time.
From this Pareto set, we choose the journey with the best utility according to the discrete choice model. 

\subsection{Parallelization of \karri}
\label{subsec:parallelization}
\begin{figure}
\FIGURE
{
    \begin{tikzpicture}[scale=1.5,
    every node/.style={align=center,scale=0.75}, 
    every path/.style={-{>[length=1.5mm,width=2mm]}}]
        \node[decision] (decmore) at (0,2.5) {more events?};
        \node[decision] (dectype) at (0,1.25) {event type?};
        \node[action] (fbi) at (0,0) {find best insertion\\for every $r\in$ batch\\in parallel};
        \node[action] (mc) at (0,-1.1) {mode choice\\with conflicts};
        \node[action] (us) at (0,-2) {update state\\in parallel};
        \node[action] (rr) at (0,-3) {remove finished\\requests};
        \node[data] (e) at (-3,2) {events};
        \node[data] (ss) at (-3,0) {system state};
        \node[data] (b) at (3,0) {batch};
        \node[decision,aspect=1.25,inner sep=0pt] (dece) at (1.75,-3) {batch\\empty?};
        \node[entry] (donerec) at (3.1,-3) {\Large $\ast$};
        \node[action] (pv) at (-3,1.25) {process vehicle\\event};
        \node[action] (ab) at (3,1.25) {add $r_i$ to\\current batch};
        \node[merge] (splitus) at (ss |- us) {};
        \node[entry] (donepv) at ([xshift=5mm,yshift=-4mm]pv.south) {\Large $\ast$};
        \node[entry] (doneab) at ([xshift=5mm,yshift=-4mm]ab.south) {\Large $\ast$};
        \node[merge, left=0.5cm of decmore] (mergeentry) {}; 
        \node[entry] at ([xshift=-0.8cm,yshift=0.2cm]mergeentry) (entry) {};
        \node[entry] at ([xshift=-0.8cm,yshift=-0.2cm]mergeentry) (entryloop) {\Large $\ast$};
        \node[exitouter, right=1cm of decmore] (exit) {};
        \node[exitinner] at (exit) {};
        \draw[] (decmore) -- node[right,pos=0.33] {yes} (dectype);
        \draw[] (decmore) -- node[above,pos=0.4] {no} (exit);
        \draw[] (dectype) -- node[above,pos=0.4] {vehicle} (pv);
        \draw[] (dectype) -- node[right,pos=0.2] {batch interval} (fbi);
        \draw[] (dectype) -- node[above,pos=0.4] {new request} (ab);
        \draw[] (fbi) -- (mc); 
        \draw[] (mc) -- (us);
        \draw[] (us) -- (rr);
        \draw[] (rr) -- (dece);
        \draw[] (dece.north) |- node[right, pos=0.05] {no} ([yshift=-3mm]fbi.east);
        \draw[] (dece) -- node[above,pos=0.33] {yes} (donerec);
        \draw[] (pv.south -| donepv.north) -- (donepv);
        \draw[] (ab.south -| doneab.north) -- (doneab);
        \draw[] (entry) to [bend left=20] (mergeentry);
        \draw[] (entryloop) to [bend right=20] (mergeentry);
        \draw[] (mergeentry) -- (decmore);
        \draw[dashed] (pv) -- (ss);
        \draw[dashed] (pv) -- (e);
        \draw[dashed] (us) -- (splitus);
        \draw[dashed] (splitus) -- (ss);
        \draw[dashed] (ss) -- (fbi);
        \draw[dashed] (b) -- (fbi);
        \draw[dashed] (ab) -- (b);
        \draw[dashed] (rr.south) -- ([yshift=-3mm]rr.south) -| ([xshift=5mm]b.east) -- (b.east);
        \draw[dashed] (splitus) -| ([xshift=-5mm]ss.west) |- (e);
        \draw[dotted,-{},thick] (b) to [bend left=20] (dece);
        \draw[dotted,-{},thick] (e) to [out=0,in=202.5] (decmore);
    \end{tikzpicture}
}{
    Workflow of \mtkarri. 
}{
    Solid arrows indicate control flow.
    Dashed arrows indicate data reads and writes.
    Dotted lines indicate information for decisions.
    The node \activityloopnode{} signifies a loop to the next event.
}{
    \label{fig:workflow_batch_parallel_karri}
}
\end{figure}
To reach the scalability required for the experiments in this paper, it does not suffice to find the best insertion and update the route state for every request one-by-one sequentially.
Therefore, we introduce a thread-based parallelization of \karri called \mtkarri.
The new algorithm computes the best insertions for a batch of requests in parallel and performs a parallelized update of the route state that incorporates the changes for the whole batch. 
We illustrate the updated workflow of the event simulation with batching and rider mode choice in~\Cref{fig:workflow_batch_parallel_karri}.

\myparagraph{Batching Requests}
To find batches of requests that can be dispatched in parallel, we partition the observation period into intervals of fixed length $\tbatch$.
The event simulation always contains a ``batch interval'' event at the end of the next interval.
When a ``new request'' event occurs, the request is not immediately dispatched but only added to the current batch.
Then, when a ``batch interval'' event occurs, the algorithm dispatches all requests in the current batch in parallel.

For this parallel process, we start one worker thread on each processing unit of the processor.
We assign each request in the batch to one worker thread that computes all shortest-path distances and the best insertion for this request, just as \karri would in the single-threaded setting.
These computations only read the current route state and distances in BCH buckets and do not change this information.
Therefore, there cannot be any data races between threads in this part.

\myparagraph{Conflicting Requests}
However, conflicts can arise if multiple requests are assigned an insertion to the same vehicle. 
Assume that both the best insertions $\ins_1$ for request $r_1$ and $\ins_2$ for request $r_2$ use vehicle $\nu$.
Performing $\ins_1$ or $\ins_2$ can change the route of $\nu$, so that the other is no longer the best insertion or even becomes infeasible.
Thus, $r_1$ and $r_2$ create a conflict on $\nu$.

\myparagraph{Iterative Conflict Resolution}
We resolve conflicts by postponing and finding new insertions for conflicting requests in an iterative manner.
Assume a conflict of $x$ requests $\{r_1, \dots, r_x\}$ where the index describes an arbitrary order of these requests.
We offer the ride resulting from the best insertion for $r_1$ to the rider.
If they reject the ride according to the mode choice model (see~\Cref{subsec:mode_choice}), then we are done with $r_1$, and we offer a ride to $r_2$, etc.
If traveler $r_i$ accepts the ride, we mark its insertion as winning and mark requests $r_{i+1}$ to $r_x$ as postponed.
After determining all winning insertions of the batch, we perform a parallelized collective update of the route state and bucket entries for these insertions.
Then, we find new best insertions for all postponed requests.
Since the insertion of $r_i$ is now represented in the route state, requests $r_{i+1}$ to $r_x$ can no longer conflict with $r_i$.
We iterate this method until all conflicts are resolved, which constitutes the inner loop at the bottom in~\Cref{fig:workflow_batch_parallel_karri}.

Choosing the batch length results in a trade-off between available parallelism and quality deterioration due to conflicts and increased wait times.
However, we find that small values of $\tbatch$ lead to a small loss in quality while providing enough parallelism for the large instances where parallelization is most important (see our experiments in~\Cref{subsubsec:tuning_batch_length}). 
Note that many different conflict resolution strategies could be implemented.
In fact, what we call a conflict here is actually an opportunity to pool requests, and a batch strategy can try to utilize synergies within a batch.
This approach has been proposed in the past as part of a rolling horizon strategy by~\cite{Kleiner2011, Agatz2011, Herbawi2012, shah_neural_2020}.
In the future, we would like to find an efficient assignment procedure that utilizes batching for better solution quality compared to our current greedy approach.

\section{Input Data and Methodology}
\label{sec:input_data_and_methodology}
In this section, we describe how we construct our problem instances and which metrics we use to judge the performance and quality of our dispatcher.

\subsection{Ride-Pooling Demand Data}
\label{subsec:ride_pooling_scenarios}
We use three different demand sources to analyze the traffic mode: The cities of Stuttgart and Karlsruhe in Germany, as well as Los Angeles in the US. 
For LA we use a public scenario~\citep{MATSimOpenLA2020} for the agent-based transport simulation MATSim~\citep{Horni2016} with different available population scales of 0.1\%, 1\%, 10\%, 25\%, 50\%  and 100\% as a baseline for the traffic demand. 
For the Karlsruhe instance, we use the demand data from \cite{worle2021modeling}, and for the Stuttgart instance we use the demand from \cite{worlen2025verkehrsentlastung}. 
Both use the agent-based simulation mobiTopp~\citep{Mallig2014}.
Each of the input data provides a mobility plan per person for the entire duration of the simulation. 
For each input plan from the simulation frameworks we extract the travel demand by extracting a start time, end time and start and end location for each mobility plan. 
For the MATSim instances we use the referenced network links as start and end coordinate and for the mobiTopp instances we use the associated coordinates from the mobility plan. 
We also re-reference person attributes to the mobility plans to add information about the agents in the simulation. 
We include information on car availability by assigning each agent a probability to be able to use a private car. 
This decision will enable or disable the option to select car as a mode for a given trip.
The mobiTopp instances allow us to derive the availability of a car for each agent from existing household information.
An agent must possess a valid drivers license to be eligible for car usage. 
All drivers in a household who have an associated personal car will have an availability of 100\%. 
The remaining drivers of the household can use a car with probability $s/u$ where $s$ is the number of shared cars and $u$ is the number of unassigned drivers.
For the LA instances, we assume that every rider has immediate access to a private car, which is not far from realistic for Los Angeles~\citep{UCDavis2024}.

\begin{figure}
\FIGURE
{    
    \input{fig/tikzplots/request_distribution_day}
}{
    Distribution of trips over a whole day in our demand sets.
}{
    Distribution of trips over a whole weekday in our demand sets Karlsruhe (Tuesday), Stuttgart (Tuesday), and Los Angeles (Wednesday).
    Shows fraction of total trips per 15-minute interval of the day.
    For Los Angeles, the plot shows the distribution for the 100\% population scale.
    The smaller scales have very similar distributions.
}{
    \label{fig:trip_distribution_day}
}
\end{figure}
We show the resulting distribution of trips over a whole weekday in~\Cref{fig:trip_distribution_day}.
For all demand sets, we see a pronounced morning and afternoon peak.
For the Karlsruhe and Stuttgart sets, the peaks are at about 07:30 am and 05:00 pm as expected.
For these demand sets, we extract only the morning peak from 06:00 to 09:00 am, which gives us the instances called \KA and \ST, respectively.
The peaks for the Los Angeles demand are notably earlier than for the other demands.
We believe that the times in the MATSim scenario use an implicit offset of three hours.
However, we could not find any documentation to this effect, so we decided to include a longer time period that covers the observed peak and the expected peak time.
Our Los Angeles instances \texttt{LA-$p$\%} for $p \in \{ 0.1, 1, 5, 10, 25, 50, 100 \}$ contain all trips that the respective $p\%$ population input files specify to be issued between 00:00 and 09:00 am.
For a summary of our problem instances, see~\Cref{tab:demand_sets}.
Note that the original demand data is based on minute intervals or quarter-hour intervals.
We describe how we convert departure times to individual seconds for our dispatcher in~\Cref{subsec:departure_times}.

\begin{table}[tb]
    \centering
    \TABLE
    {Overview on demand sets}
    {\begin{tabular}{@{}
    l
    c
    S[table-format = 8, table-alignment-mode = format, table-number-alignment=center]
    r
    S[table-format = 3.2, table-alignment-mode = format, table-number-alignment=center]
    c
    @{}}
        \toprule
        Name  &  Network & \multicolumn{1}{c}{\makecell[c]{\#trips}} & \makecell[c]{period} & \multicolumn{1}{c}{\makecell[c]{\#req./$\s$}} & source  \\
        \midrule
        \KA             & Karlsruhe   &   975773 & 6:00-9:00 &  90.35 & mobiTopp \\
        \ST             & Stuttgart   &  1388868 & 6:00-9:00 & 128.60 & mobiTopp \\
        \LAzero         & Los Angeles &    25415 & 0:00-9:00 &   0.78 & MATSim \\
        \LAone          & Los Angeles &   254203 & 0:00-9:00 &   7.85 & MATSim \\
        \LAfive         & Los Angeles &  1270531 & 0:00-9:00 &  39.21 & MATSim \\
        \LAten          & Los Angeles &  2544143 & 0:00-9:00 &  78.52 & MATSim \\
        \LAtwentyfive   & Los Angeles &  6359125 & 0:00-9:00 & 196.27 & MATSim \\
        \LAfifty        & Los Angeles & 12719226 & 0:00-9:00 & 392.57 & MATSim \\
        \LAhundred      & Los Angeles & 25431754 & 0:00-9:00 & 784.93 & MATSim \\
        \bottomrule
    \end{tabular}}{}{
    \label{tab:demand_sets}
    }
\end{table}

\myparagraph{Walking Speed and Walking Radius based on Rider Properties}
We determine the maximum walking radius by taking the comfortable walking speed as measured by \cite{bohannon1997comfortable}. To convert the walking speed into a distance we multiply the speed with a fixed duration.
We determine this duration by taking the access penalty parameter of ridesharing from the utility calculation (see~\Cref{subsec:mode_choice}) and set a fixed maximum so that the penalty from the access time is causing the utility to drop at most to 50\% of the original utility. The resulting accepted walking distances are around 200--250 meters, which is within the recommended transit stop density of 300 meters for a high quality service area \citep{BBSR2018_OeV_Angebotsqualitaet}. 

\myparagraph{Vehicle and Pedestrian Networks}
\begin{table}[t!]
\TABLE
{
    Overview on road networks
}{
    \begin{tabular}{@{}
    c
    S[table-format = 6, table-alignment-mode = format, table-number-alignment=center]
    S[table-format = 7, table-alignment-mode = format, table-number-alignment=center]
    c
    @{}}
        \toprule
        Name  & \multicolumn{1}{c}{\makecell[c]{$|V|$}} & \multicolumn{1}{c}{\makecell[c]{$|E|$}} & Network Structure \\
        \midrule
        Karlsruhe & 276314 & 574514 & single center \\ 
        Stuttgart & 352855 & 758452 & single center \\ 
        Los Angeles & 559676 & 1220393 & no strong center \\
        \bottomrule
    \end{tabular}
}{}{
    \label{tab:networks}
}
\end{table}
We generate road networks based on OpenStreetMap data (see~\Cref{subsec:generating_road_networks_and_determining_meeting_points} for details).
We specify key metrics of the networks in~\Cref{tab:networks}.

\subsection{Definition of Urban Core and Periphery}
\label{subsec:definition_core_periphery}
\begin{figure}[t]
\FIGURE
{
    \subcaptionbox
    {Germany\label{fig:urban_rural_map_germany}}
    {\includegraphics[width=.3\textwidth, height=5cm, keepaspectratio]{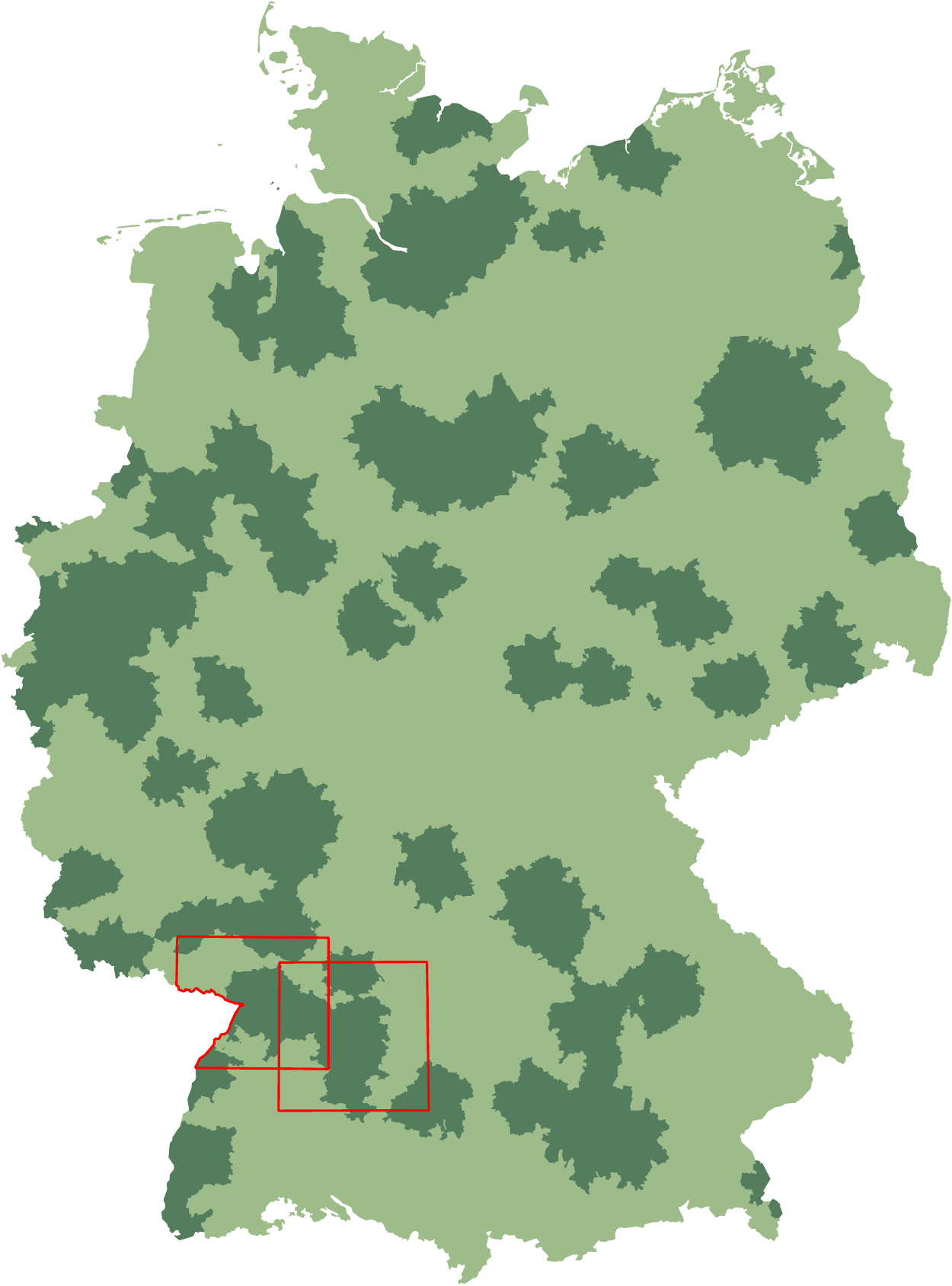}}%
    \hfill\subcaptionbox
    {Stuttgart\label{fig:urban_rural_map_stuttgart}}
    {\includegraphics[width=.3\textwidth, height=4cm, keepaspectratio]{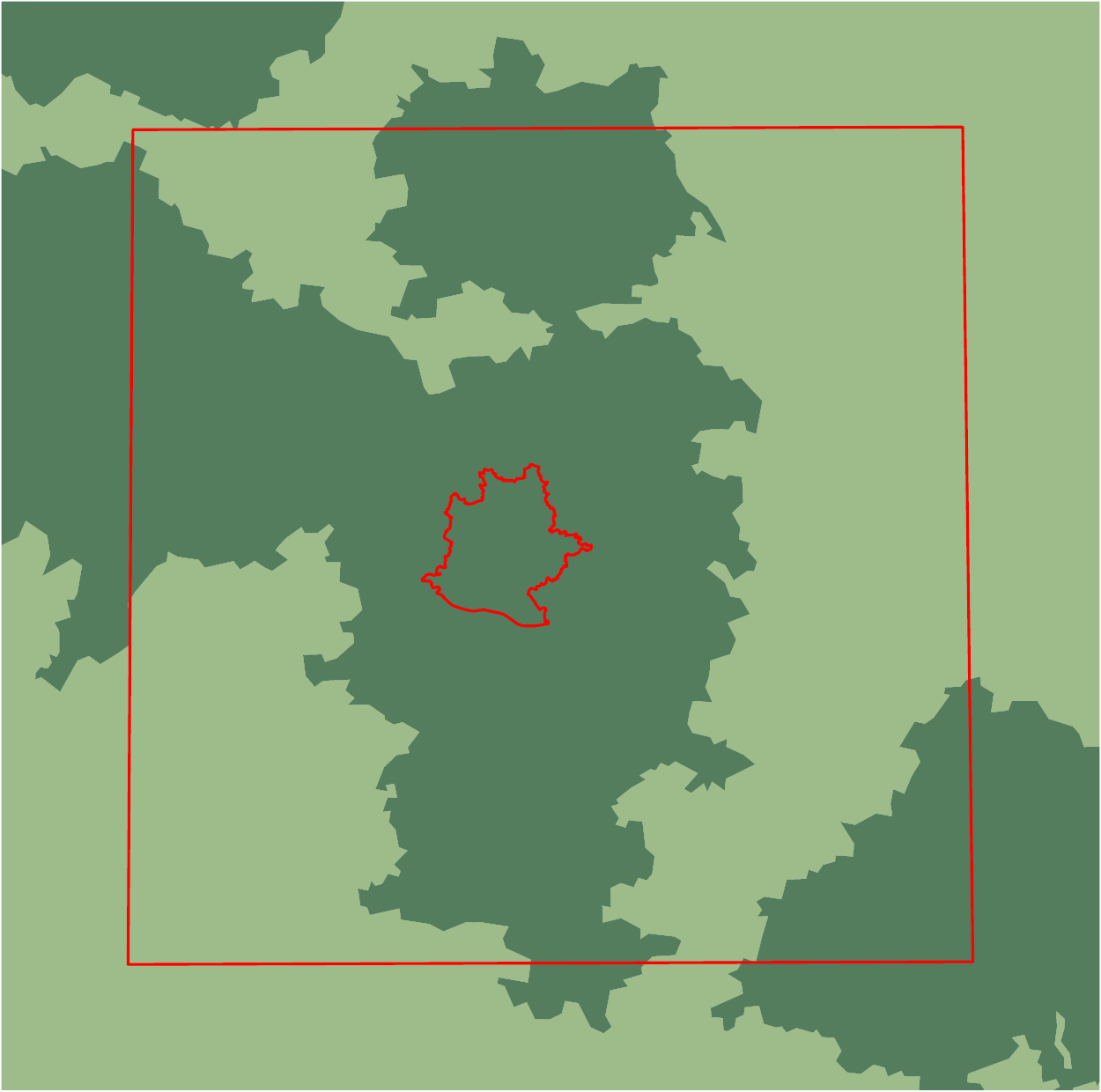}}%
    \hfill\subcaptionbox
    {Karlsruhe\label{fig:urban_rural_map_karlsruhe}}
    {\includegraphics[width=.3\textwidth, height=4cm, keepaspectratio]{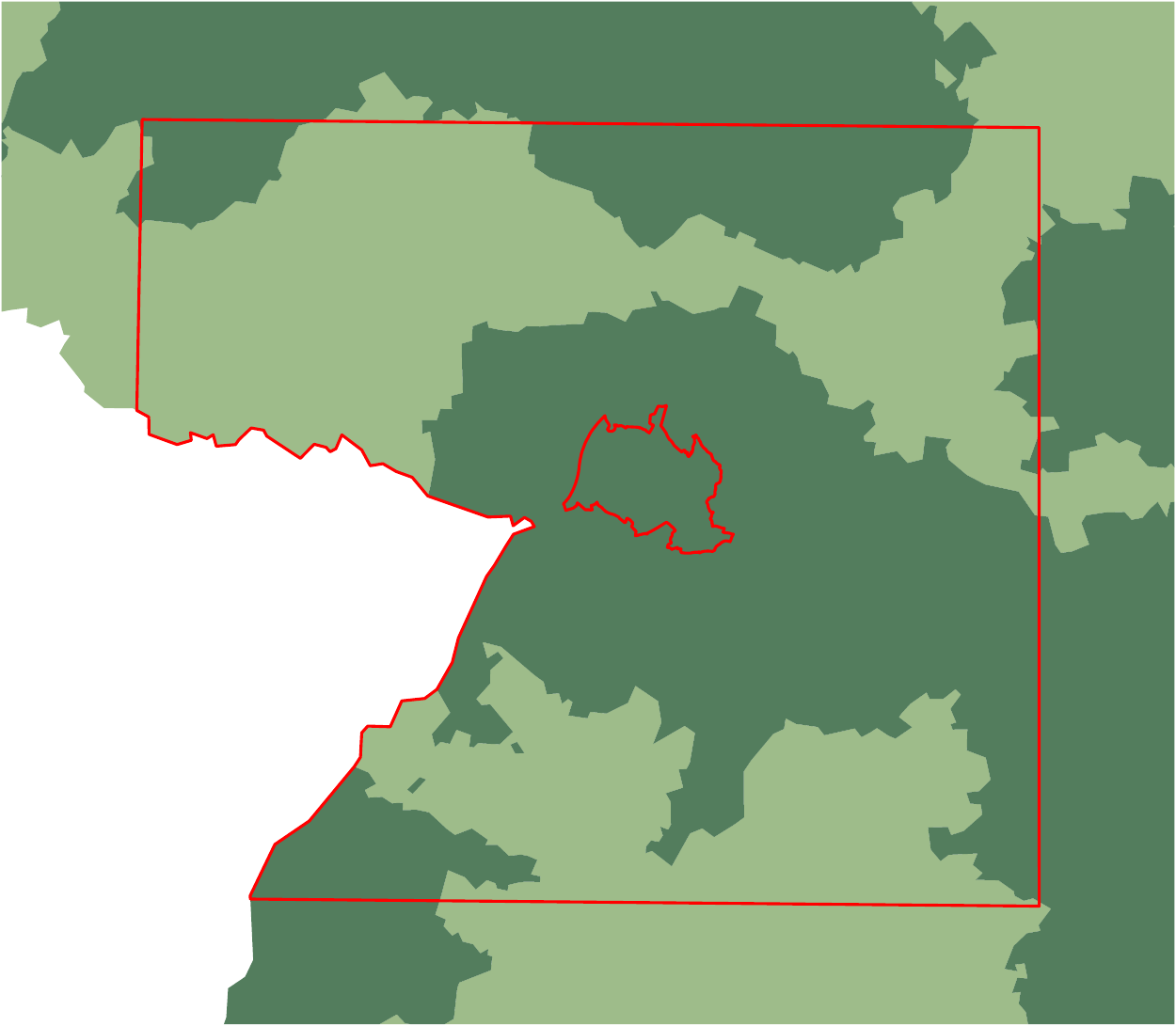}}
}{
    RegioStaR-2 typology of Germany, Stuttgart and Karlsruhe.
}{
    Shows entire country of Germany (\Cref{fig:urban_rural_map_germany}) and the areas immediately surrounding Stuttgart (\Cref{fig:urban_rural_map_stuttgart}) and Karlsruhe (\Cref{fig:urban_rural_map_karlsruhe}).
    Urban areas are shaded dark, rural areas are shaded light.
    Red boxes in~\Cref{fig:urban_rural_map_germany} show position of Stuttgart (right box) and Karlsruhe (left box) within Germany.
    For~\Cref{fig:urban_rural_map_stuttgart,fig:urban_rural_map_karlsruhe}, the inner red outline indicates the limits of the city proper and the outer red outline indicates the area in which trips are considered.
    The white area in the left of~\Cref{fig:urban_rural_map_karlsruhe} are parts of France for which we do not have RegioStaR data.
    No trips are made in this area.
}{
  \label{fig:urban_rural_map}
}
\end{figure}
We partition the spatial topology of the German scenario instances based on the official German RegioStaR typology \citep{BMDV2018RegioStaR}. 
This typology categorizes the municipalities based on settlement structure, centrality and spatial location. 
The typology can be differentiated at varying levels of precision to incorporate more information about the municipalities. 
We use the coarsest level RegioStaR-2, which separates between urban and rural regions. 
\Cref{fig:urban_rural_map} visualizes the RegioStaR-2 classification for all of Germany and for the areas immediately surrounding the cities of Stuttgart and Karlsruhe.
Note that the area surrounding Karlsruhe contains parts of France for which we do not have a categorization.
The \KA demand set is limited to trips that lie entirely within Germany, though.

\subsection{Performance and Quality Metrics}
\label{subsec:metrics}
We measure the performance of the dispatcher by running time per request and the speedups observed by using multiple processors. 
To capture the system usage we measure the absolute number of ride-pooling users (\emph{\# riders})  and their relative share of total trips (\emph{RP modal share}) by dividing the number of ridepooling trips by the number of total trips. We evaluate system-level efficiency by comparing the total time vehicles spend on the road to a baseline in which each ride-pooling user instead used their own car. 
Let $T_\text{car}$ be the total driving time of private cars if every ride-pooling user used an individual car instead.
Let $T_\text{RP}$ be the total driving time of ride-pooling vehicles.
We report the ratio $T_\text{car} / T_\text{RP}$ (\emph{system effectiveness}).
A system effectiveness of $x$ means that the total vehicle operation time needed for all riders is reduced by a factor of $x$ by using ride-pooling instead of private cars.
We evaluate vehicle utilization using the time-weighted average occupancy (\emph{occupancy}), which includes any time periods in which a vehicle is moving.

To assess waiting times, we report the average absolute waiting time of all ride-pooling trips (\emph{wait time}).
For the travel time, we compute the detour experienced by each rider, i.e. the difference between the in-vehicle time of the ride-pooling trip and the direct travel time by car.
Let $\delta_\text{car}$ be the direct travel time from a rider's origin to their destination.
Let $\delta_\text{RP}$ be the total time the rider spends in a ride-pooling vehicle.
We report the average of the ratios $\delta_\text{RP} / \delta_\text{car}$ for all riders (\emph{ride detour}).
In addition, we compute the average number of co-riders (\emph{number of co-riders}) per ride-pooling user as a measure of the degree of pooling experienced by passengers.

\section{Experimental Results}
\label{sec:experimental_results}
In this section, we describe the results of our experimental simulation study.
First, we analyze the performance of our dispatcher to demonstrate its suitability for simulations at a huge scale.
Then, we analyze the solution quality of the dispatcher for varying configurations of the problem instances. 

\subsection{Setup}
\label{subsec:experimental_setup}
Our source code (available at~\url{https://github.com/molaupi/karri}) is written in C++ 20 and compiled with GCC 11.5 using \texttt{-O3}.
We use multi-threading primitives from the Intel OneAPI Threading Building Blocks library (version 2021.13.1).
We run our experiments on two separate machines that both run Rocky Linux 9.4.
Our experiments on running times use a machine with an AMD EPYC 9684X processor (96 physical cores) and 1.6 TB of RAM. 
Since hardware has no effect on solution quality, we use a machine with an AMD EPYC 7702 processor (64 physical cores) and 1.0 TB of RAM for some quality experiments. 
Note that even on our largest instances, we actually need only 60 GB of RAM.
We use 32-bit distance labels, so \karri's vectorized searches using the AVX2 SIMD instruction set with 256-bit registers allow for eight simultaneous searches.

Unless otherwise specified, we use the following default assumptions:
We use the model parameters decided on in the parameter tuning experiments in the appendix (\Cref{sec:tuning_of_karri_model_parameters}) and $\tbatch=5\s$.
Vehicles have a default capacity of four, and the initial location of every vehicle is an edge drawn uniformly at random in the road network of the service area.
The service time of every vehicle is set to encompass the entire observation period.
We use all threads available on the machine.
We specify the number of runs per experiment and report average metric values over all runs.

\subsection{Parallelization and Scalability}
\label{subsec:parallelization_and_scalability_experiments}
In this section, we evaluate our parallelization and the scalability of \mtkarri.

\subsubsection{Tuning $\tbatch$}
\label{subsubsec:tuning_batch_length}
We begin by analyzing the effect of the batch length parameter $\tbatch$ on the solution quality and the run time of \mtkarri compared to sequential \karri.

\myparagraph{Effect of $\tbatch$ on Solution Quality}
\begin{figure}
\FIGURE
{    
    \input{fig/tikzplots/combined_batch_length_quality_plot}
}{
    Effect of batch length on solution quality, relative to non-batched \karri.
}{
    Change in quality for different $\tbatch$ compared to the non-batched version of \karri on \KA and \LAtwentyfive with \num{50000} and \num{125000} vehicles, respectively.
    Shows the change for RP modal share, occupancy, and number of co-riders, as well as ride-detour, wait time, and system effectiveness.
    For explanations of the metrics, see~\Cref{subsec:metrics}.
    Note the logarithmic x-axis.
}{
  \label{fig:quality_batch_lengths}
}
\end{figure}
We compare the quality of assignments produced by our parallel implementation using different values of $\tbatch$ to the quality of the original sequential code.
For this, we show the quality change for different values of $\tbatch$ compared to the non-batched version for the \KA and \LAtwentyfive instances in~\Cref{fig:quality_batch_lengths}.
The plots show averages of three runs. 

The values of most metrics exhibit a linear relationship with the length of batch intervals (note the logarithmic x-axis).
The waiting time increases the most, since a traveler may have to wait up to $\tbatch$ to be processed as part of the next batch.
This also causes a slight drop in the modal share, occupancy, and number of co-riders for larger values of $\tbatch$.
Noticeably, even at $\tbatch=1\s$, there is a negative offset in pooling (number of co-riders) compared to the non-batched version, which is more pronounced for the larger \LAtwentyfive instance. 
This can be attributed to the order in which requests are processed.
Consider the latest request $r$ of a batch.
In the sequential setting, all other requests in the batch have already been processed and are candidates for a pooled ride of $r$.
In the batched setting, the other requests have not been processed yet, and a synergy between $r$ and the other requests may be overlooked.
Thus, in the batched setting, there are always slightly more non-pooled rides than without batches.

\myparagraph{Effect of $\tbatch$ on Running Time}
\begin{figure}
\FIGURE
{
    \input{fig/tikzplots/batch_length_speedups}
}{
    Effect of batch length on speedup over sequential \karri.
}{
    Average speedup of \mtkarri with $T=96$ threads over sequential \karri for batch lengths $\tbatch \in \{1,2,5,15,30\}$.
    The \texttt{LA-$p$\%} instances use $p \cdot 5000$ vehicles, and the \KA and \ST instances use \num{50000} vehicles.
}{
  \label{fig:running_time_batch_lengths}
}
\end{figure}
We now evaluate how the value of $\tbatch$ affects the running time of the batch-parallel dispatcher.
For this, we show the speedup over sequential \karri for $T=96$ threads and varying $\tbatch$ in~\Cref{fig:running_time_batch_lengths}.
We report averages of five runs for instances up to \ST, averages of three runs for \LAtwentyfive, and results for a single run for the largest instances \LAfifty and \LAhundred.
As sequential \karri would have infeasible running times on large instances, we obtain a sample of the running times of \karri by running the parallel \mtkarri and solving every one-hundredth request using the sequential algorithm.
The speedups for the large instances (\LAtwentyfive and larger) use this sample as the sequential baseline.

We observe that even small values of $\tbatch$ suffice for good speedups with $T=96$ threads.
For the largest instances (\LAfifty and \LAhundred), batches are already large enough at $\tbatch=5\s$ to provide sufficient parallelism, with speedups of \num{36} or larger.  
Beyond $\tbatch=5\s$ speedups increase only slightly for large instances.
For smaller instances, speedups are worse due to limited parallelism available in every batch.
This is not a drawback though, as smaller instances already have shorter total running times.
Since there is a trade-off with quality, we choose to use $\tbatch=5\s$ as the default value for fast and high-quality parallel dispatching.

\subsubsection{Scalability of Simulation}
\label{subsubsec:scalability}
In this section, we experimentally evaluate the scalability of \mtkarri to show that our dispatcher can solve large instances quickly.
To judge the performance of parallel code, we can consider \emph{strong scaling} or \emph{weak scaling}.
Strong scaling describes the reduction in running times for an increasing number of processing units and a fixed \emph{total} problem size.
Weak scaling describes how the running time varies for an increasing number of processing units and a fixed problem size \emph{per processing unit}.
We focus on weak scaling here and show strong scaling experiments in~\Cref{sec:strong_scaling}.
We conduct only a single run of experiments in this section, since scaling experiments take a long time, and one run suffices to see the trend in running times.

\myparagraph{Weak Scaling}
\begin{table}[t]
\TABLE
{
    Weak scaling of \mtkarri.
}{
    \begin{tabular}{@{}
    l
    S[table-format = 2, table-alignment-mode = format, table-number-alignment=center]
    S[table-format = 5, table-alignment-mode = format, table-number-alignment=center]
    S[table-format = 4, table-alignment-mode = format, table-number-alignment=center]
    S[table-format = 2.1, table-alignment-mode = format, table-number-alignment = center]
    S[table-format = 2.1, table-alignment-mode = format, table-number-alignment = center]
    S[table-format = 2.1, table-alignment-mode = format, table-number-alignment = center]
    S[table-format = 2.2, table-alignment-mode = format, table-number-alignment=center]
    S[table-format = 1.2, table-alignment-mode = format, table-number-alignment=center]
    @{}}
    \toprule
         \makecell[c]{instance} & $T$ & \multicolumn{1}{c}{\makecell[c]{\# req.\\{}\scriptsize [$\cdot 10^3$]}} & \multicolumn{1}{c}{\makecell[c]{$|F|$\\{}\scriptsize [$\cdot 10^3$]}} & \multicolumn{1}{c}{\makecell[c]{find ins.\\{}\scriptsize (speedup)}} & \multicolumn{1}{c}{\makecell[c]{update\\{}\scriptsize (speedup)}} & \multicolumn{1}{c}{\makecell[c]{total\\{}\scriptsize (speedup)}} & \multicolumn{1}{c}{\makecell[c]{$t_\text{seq}$\\{}\scriptsize [$\ms$]}} & \multicolumn{1}{c}{\makecell[c]{$t_\text{par}$\\{}\scriptsize [$\ms$]}} \\        
         \cmidrule(r){1-4}\cmidrule(lr){5-7}\cmidrule(l){8-9}
         \LAone         &  1 &   254 &   5 &  0.97 &  0.92 &  0.96 &  0.65 & 0.68 \\
         \LAfive        &  5 &  1271 &  25 &  4.32 &  3.71 &  4.21 &  2.31 & 0.55 \\
         \LAten         & 10 &  2544 &  50 &  8.21 &  4.60 &  7.65 &  4.58 & 0.60 \\
         \LAtwentyfive  & 25 &  6359 & 125 & 16.92 &  8.37 & 15.68 & 12.60 & 0.80 \\
         \LAfifty       & 50 & 12719 & 250 & 29.10 &  9.32 & 26.71 & 30.68 & 1.15 \\
         \LAhundred     & 96 & 25432 & 500 & 39.43 & 13.69 & 36.89 & 68.00 & 1.84 \\
         \midrule
         \LAhundred     & 96 & 25432 &   5 & 70.84 &  5.14 & 49.72 &  2.06 & 0.04 \\
         \LAhundred     & 96 & 25432 &  50 & 60.54 & 10.62 & 53.65 & 19.00 & 0.35 \\
    \bottomrule
    \end{tabular}
}{
    Speedup of \mtkarri over sequential \karri with proportionally increasing instance size ($\#$ req. and $|F|$) and number of threads ($T$), plus two additional configurations for the largest instance.
    Shows speedups for finding ride-pooling insertions (find ins.), updating the system state for accepted rides (update), and the total running time (total).
    Additionally shows average running time per request (in $\ms$) for \karri ($t_\text{seq}$) and \mtkarri ($t_\text{par}$). 
}{
    \label{tab:speedups_weak_scaling}
}
\end{table}
We run sequential \karri and parallel \mtkarri on \texttt{LA-$p$\%} for $p \in \{ 1, 5, 10, 25, 50, 100 \}$ to scale the number of requests proportionally to $p$. 
We also increase the number of vehicles proportionally.
For \mtkarri, we start with one thread for the \LAone instance and increase the number of threads in proportion to $p$.
We report the speedups of \mtkarri over sequential \karri in~\Cref{tab:speedups_weak_scaling}.
Remark two caveats:
First, we use \num{96} threads for the \LAhundred instance, since the machine used in the experiments only has \num{96} cores.
Second, the sequential baseline for large instances (\LAfifty and \LAhundred) uses only a sample of requests as in~\Cref{subsubsec:tuning_batch_length}.

We see good speedups for \texttt{LA-$p$\%} with $p$ threads for $p \le 10$.
Here, the efficiency (speedup divided by number of threads) is at least \num{0.82} for finding the best insertions.
The work for updating the system state is not parallelized well and the efficiency is much lower.
As most of the time is spent finding insertions, we still see good speedups of the total running time (efficiency $\ge$ \num{0.76}). 

The scalability appears to reach a limit for larger instances, with speedups of the total running time of only \num{15.68}, \num{26.71} and \num{36.89} with \num{25}, \num{50} and \num{96} threads, respectively.
For an explanation, we see multiple contributing factors.
Firstly, since we use the same batch length of $\tbatch=5\s$ for all instances, the number of requests per batch increases proportionally to $p$ for the \texttt{LA-$p\%$} instances.
Thus, there are more conflicting insertions per batch, which leads to additional iterations in our conflict resolution approach and more re-computations of insertions.
We can see the extent of this effect by normalizing running times with the total number of insertions computed instead of the number of requests.
Even then, the speedups for \LAfifty and \LAhundred would be only \num{35.77} and \num{53.05}.
This brings us to the second reason for the limited scalability.
Consider the running time per request for \karri ($t_\text{seq}$) and \mtkarri ($t_\text{par}$) given in~\Cref{tab:speedups_weak_scaling}.
For the sequential algorithm, the running time per request increases in proportion to the size of the instance. 
This is because there are more vehicles and more riders, which means that the BCH buckets have more entries and each shortest-path query becomes more complex.
For the parallel algorithm, the increased number of threads is not able to compensate for this effect as we see an increasing running time per request with larger instances.
We believe that the parallel algorithm is affected more by the increased complexity of the BCH data structure because the uncoordinated concurrent reads of the buckets lead to bad caching.
To gain an understanding of the speedups possible without this drawback, we include the last two rows in~\Cref{tab:speedups_weak_scaling} where we consider \LAhundred with smaller fleets.
Although the speedups for updating the system state are still limited, we see much better speedups for finding insertions of \num{70.84} for \num{5000} vehicles and \num{60.54} for \num{50000} vehicles, which leads to total speedups of \num{49.72} and \num{53.65}.
In~\Cref{sec:strong_scaling}, we support this with experiments on strong scaling.

Thus, a potential avenue for optimization is an improved BCH data structure and more coordinated access to this data structure to improve locality.
For example, requests within the same batch may be ordered so that requests with origin and destination locations that are close to each other are processed at the same time.
Due to this proximity in the road network, the BCH queries for these requests would access similar buckets, improving the locality of the memory reads.

\subsection{Quality}
In this section, we use \mtkarri to analyze the quality of ride-pooling on large metropolitan road networks with respect to rider service quality and vehicle resource usage.

\subsubsection{Impact of Customer Base}
We start by analyzing how the size of the customer base, i.e. the share of the population that considers using ride-pooling, affects the quality of the ride-pooling system.
We use the \texttt{LA-$p$\%} input instances (for $p \in \{ 0.1, 1, 5, 10, 25, 50, 100 \}$, s.~\Cref{subsec:ride_pooling_scenarios}) with fixed fleet sizes to represent different portions of the population making queries to our dispatcher.
We report averages of three runs.
Note that we have only limited information on individual travelers for the \texttt{LA} instances.
Therefore, in this section, we do not use meeting points, since we don't know walking preferences of travelers.

\myparagraph{Number of Riders}
\begin{figure}[tb]
\FIGURE
{
    \input{fig/tikzplots/LA_riders_plots}
}{
    Effect of customer base on ridership.
}{
    Ridership statistics for the \texttt{LA-$p$\%} instances with $p \in \{ 0.1, 1, 5, 10, 25, 50, 100 \}$ using different fleet sizes.
    The x-axis describes the share of the population willing to use ride-pooling on a logarithmic scale.
    The y-axis of each plot describes the number of riders in millions (left) and the RP modal share (right).
    Note the different y-axis scales.
    For explanations of the metrics, see~\Cref{subsec:metrics}.
}{
  \label{fig:request_density_riders_plot_la}
}
\end{figure}
The plots in~\Cref{fig:request_density_riders_plot_la} show the total number of riders as well as the ratio of travelers choosing ride-pooling over other modes of transport for a growing customer base.

The number of riders increases monotonically with a growing customer base.
For the largest fleet, there are more than nine million total riders.
For fleets of \num{50000} or fewer vehicles, the number of riders increases only slowly.
The RP modal share remains high for large fleets (up to \num{37}\% of the total population choose ride-pooling) but drops sharply for small fleets and a large customer base.

\myparagraph{Pooling}
\begin{figure}[tb]
\FIGURE
{
    \input{fig/tikzplots/LA_sharing_plots}
}{
    Effect of customer base on pooling.
}{
    Pooling metrics on the \texttt{LA-$p$\%} instances with $p \in \{ 0.1, 1, 5, 10, 25, 50, 100 \}$ using different fleet sizes.
    The x-axis describes the share of the population willing to use ride-pooling on a logarithmic scale.
    The y-axis shows the number of co-riders (left) and occupancy (right).
    Note the different y-axis scales.
    For explanations of the metrics, see~\Cref{subsec:metrics}.
}{
  \label{fig:request_density_sharing_plot_la}
}
\end{figure}
In~\Cref{fig:request_density_sharing_plot_la}, we show plots for three metrics describing the extent of shared rides in the ride-pooling system.
First, we observe that the average number of co-riders increases with a larger customer base.
This effect is relevant even for the largest vehicle fleets considered.
Thus, a high rate of accepted rides and a large portion of shared rides are not mutually exclusive. 
For example, more than five million travelers choose ride-pooling for the largest customer base and \num{125000} vehicles. 
At the same time, every rider has an average of \num{1.79} co-riders along their trip.

This trend of increased pooling is also reflected in the average vehicle occupancy.
For a small customer base, the occupancy is below one because of dead mileage spent moving to a request's origin.
However, for large customer bases, this effect is compensated for by improved pooling.
For \num{125000} vehicles and the largest customer base, an average of \num{2.32} riders occupy each vehicle while driving.
Even for the largest vehicle fleet considered, there are about \num{1.45} riders on average.

\myparagraph{Quality of Rides, Resource Usage}
\begin{figure}[tb]
\FIGURE
{
    \input{fig/tikzplots/LA_quality_plots}
}{
    Effect of customer base on trip quality and vehicle resource usage.
}{
    Quality metrics on the \texttt{LA-$p$\%} instances with $p \in \{ 0.1, 1, 5, 10, 25, 50, 100 \}$ using different fleet sizes.
    The x-axis describes the share of the population willing to use ride-pooling on a logarithmic scale.
    The y-axis of each plot shows the ride detour, wait time, and system effectiveness, respectively.
    Note the different y-axis scales.
    For an explanation of the metrics, see~\Cref{subsec:metrics}.
}{
  \label{fig:request_density_quality_plot_la}
}
\end{figure}
The plots in~\Cref{fig:request_density_quality_plot_la} show the quality of accepted ride-pooling rides and the usage of vehicle resources relative to the size of the customer base.

We observe growing ride detours for larger customer bases.
This is caused by the previously noted higher level of pooling and the detours made for shared rides.
Since smaller fleets require more pooling than large ones, small fleets also lead to larger in-vehicle times.
With \num{500000} vehicles, the average in-vehicle time is at most \num{18}\% longer than a direct car trip, but with \num{125000} vehicles, the average in-vehicle time is up to \num{68}\% longer.
Similarly, wait times increase with larger customer bases because vehicles may have to make stops for other riders before picking up a new rider.
Due to our randomized mode choice model, even some of the rides with very long wait times are accepted.

Finally, we consider the system effectiveness as a measure of the resources saved compared to private cars.
With small customer bases, there is little pooling, so the dead mileage spent to pick up new riders dominates the driving time saved through pooling. 
Here, ride-pooling can be assumed to use more vehicle resources than private cars (system effectiveness smaller than one).
However, the improved pooling that comes with a larger customer base inverts this ratio such that ride-pooling vehicles spend less time driving than private cars would.
For large vehicle fleets, ride-pooling uses up to \num{1.57} times less total vehicle time on the road than private cars.

\myparagraph{Conclusion on Customer Base}
As expected, we find that larger customer bases increase the number of travelers who use ride-pooling and can increase the potential for shared rides, which leads to significant savings in vehicle resources.
However, the quality of rides deteriorates significantly if a small fleet tries to handle a high demand for rides.
In these situations, the dispatcher should potentially make the decision to no longer offer new rides to avoid this degradation of quality.

\subsubsection{Impact of Meeting Points}
\label{subsubsec:impact_of_meeting_points}
\begin{table}
\TABLE
{
    Effect of meeting points on quality of trips and usage of vehicle resources.
}{
    \setlength\tabcolsep{4pt}
    \begin{tabular}{@{}
    c
    c
    S[table-format = 2.2, table-alignment-mode = format, table-number-alignment=center]
    S[table-format = 2.2, table-alignment-mode = format, table-number-alignment=center]
    r
    r
    r
    r
    r
    S[table-format = 2.2, table-alignment-mode = format, table-number-alignment=center]
    S[table-format = 5, table-alignment-mode = format, table-number-alignment=center]
    S[table-format = 2.2, table-alignment-mode = format, table-number-alignment=center]
    @{}}
    \toprule
        inst. & walk? & \multicolumn{1}{c}{$|P|$} & \multicolumn{1}{c}{$|D|$} & \makecell[c]{p-walk\\{}\scriptsize[mm:ss]} & \makecell[c]{d-walk\\{}\scriptsize[mm:ss]} & \makecell[c]{wait\\{}\scriptsize[mm:ss]} & \makecell[c]{in-veh.\\{}\scriptsize[mm:ss]} & \makecell[c]{trip\\{}\scriptsize[mm:ss]} & \multicolumn{1}{c}{\makecell[c]{co-riders\\{}\scriptsize[abs.]}} & \multicolumn{1}{c}{\makecell[c]{drive\\{}\scriptsize[h]}} & \multicolumn{1}{c}{\makecell[c]{RP\\{}\scriptsize[\%]}} \\
    \midrule
        \multirow{3}{*}{\KA} & no & 1.00 &  1.00 & 00:00 & 00:00 & 05:46 & 18:21 & 24:07 & 1.50 & 55093 & 40.85 \\
        & yes & 14.92 & 14.85 & 00:53 & 00:12 & 04:00 & 17:06 & 22:11 & 1.57 & 49868 & 41.09 \\
        \cmidrule{2-12}
        & rel [\%] &  &  &  &  & -30.63 & -6.85 & -8.01 & 4.32 & \multicolumn{1}{r}{-9.48} & 0.60 \\
    \midrule
        \multirow{3}{*}{\ST} & no & 1.00 &  1.00 & 00:00 & 00:00 & 16:02 & 28:57 & 44:58 & 1.96 & 75786 & 30.77 \\
        & yes & 19.04 & 16.10 & 01:18 & 00:21 & 11:44 & 26:36 & 39:58 & 2.06 & 70416 & 32.52 \\
        \cmidrule{2-12}
        & rel [\%] &  &  &  &  & -26.85 & -8.10 & -11.12 & 5.00 & \multicolumn{1}{r}{-7.08} & 5.70 \\
    \bottomrule
    \end{tabular}
}{
    Effect of meeting points with walking on dispatcher quality for the \KA and \ST instances with \num{50000} vehicles.
    Shows average number of pickup locations (\textit{$|P|$}) and dropoff locations (\textit{$|D|$}) considered, average pickup walking time (\textit{p-walk}), dropoff walking time (\textit{d-walk}), wait time (\textit{wait}), in-vehicle time (\textit{in-veh.}) and total trip time (\textit{trip}) of riders, as well as average number of co-riders along each rider's trip (\textit{co-riders}), total vehicle drive time (\textit{drive}), and RP modal share (\textit{RP}).
    Row marked ``no'' shows results without meeting points.
    Row marked ``yes'' shows results with traveler-specific meeting points determined using walking speed and maximum walking distance defined for each individual traveler (see~\Cref{subsec:ride_pooling_scenarios}).
    Rows marked ``rel. [\%]'' give change relative to no walking in percent.
}{
    \label{tab:walking_quality}
}
\end{table}
For the rest of the experimental section, we use our dispatcher with meeting points, i.e. riders walking to pickup locations and from dropoff locations (see~\Cref{subsec:ride_pooling_model}). 
We motivate this using the following experiments that show the benefits of meeting points for the solution quality of our ride-pooling dispatcher.
The set of meeting points for each rider is generated as described in~\Cref{subsec:generating_road_networks_and_determining_meeting_points}.
\Cref{tab:walking_quality} shows the quality without and with meeting points for the \KA and \ST instances.
We report average results for five runs of each experiment.

Given traveler-specific walking speeds and distances, an average of between \num{15} and \num{19} pickup and dropoff locations are considered for each request.
On average, each rider walks about five times farther in the beginning of their trip than in the end because vehicles are more likely to be constrained by existing riders at the pickup location than at the dropoff location.
For both instances, the wait time shrinks considerably when using meeting points.
In the \ST instance, riders wait a long time, but meeting points reduce the average wait time by more than four minutes. 
The average time between issuing a request and being picked up is reduced from a pure waiting time of 05:46 min (16:02 min) to a combination of walking and waiting comprising 05:05 min (12:05 min) for the \KA (\ST) instance.
Additionally, the average in-vehicle time for riders decreases by \num{6.85}\% for \KA and \num{8.10}\% for \ST, leading to average total trip times that are \num{8.01}\% and \num{11.12}\% smaller than without meeting points for \KA and \ST.
The level of pooling improves when using meeting points, with the average number of co-riders increasing by \num{4.32}\% for the \KA instance and \num{5.00}\% for the \ST instance.
This is also reflected in a decrease in total vehicle drive time by \num{9.48}\% for \KA and \num{7.08}\% for \ST.

The modal share of ride-pooling stays about the same for the \KA instance, but increases by \num{5.70}\% for the \ST instance.
This difference between the instances is likely due to the fact that the \KA instance has fewer requests and a smaller network.
Thus, since we use fleets of the same size for both instances, the fleet is under more stress in the \ST instance and the quality of rides is worse.
Therefore, for the \ST instance, meeting points have a greater impact on the quality of ride-pooling compared to other modes of transportation, which leads to a greater improvement in modal share.

In conclusion, meeting points improve both rider trips and vehicle resource usage.
They are particularly helpful if the system is experiencing high demand.
Thus, we always consider meeting points if possible.
In a real-world system, it might make sense to offer both a ride with walking and a ride without walking and include incentives for walking (e.g. an additional fare discount).

\subsubsection{Impact of Fleet}
\label{subsubsec:impact_of_fleet}
\begin{figure}[tb]
\FIGURE
{
    \input{fig/tikzplots/Karlsruhe_fleet_quality_plot}
}{
    Effect of fleet size on quality.
}{
    Quality on \KA with fleets $F$ of $|F| \in \{ \num{10}, \num{20}, \num{30},\num{40},\num{50},\num{75},\num{100}, \num{150}, \num{200}, \num{300}, \num{400}, \num{500} \} \cdot 10^3$ vehicles and $\capacity(\nu) \in \{2,4,8\}$ for every $\nu \in F$ with and without meeting points (\textit{MPs}). 
    The x-axis describes the number of vehicles in thousands. 
    The y-axis shows (left to right, top to bottom) RP modal share, ride detour, average wait time, occupancy, number of co-riders, and system effectiveness, respectively.
    For explanations of the metrics, see~\Cref{subsec:metrics}.
    Note the different y-axis scales.
}{
  \label{fig:fleet_quality_plot_karlsruhe}
}
\end{figure}
 
\Cref{fig:fleet_quality_plot_karlsruhe} depicts plots on the solution quality relative to the size of the fleet and vehicle seating capacity for the \KA instance.
The same plot for \ST can be found in~\Cref{fig:fleet_quality_plot_stuttgart} in~\Cref{sec:omitted_plots}.
We report average results for three runs.

First, notice that the use of meeting points leads to improved quality in all metrics. 
Although the modal share of ride-pooling is not affected much, in-vehicle times and wait times become shorter, pooling is improved, and the system effectiveness is around \num{10}\% better than without meeting points, irrespective of the size of the fleet.
Thus, we focus on the case with meeting points.

We find that the share of travelers using ride-pooling increases with a growing number of vehicles.
This can be attributed to an improved quality of rides as evidenced by smaller ride detours and wait times.
Until about \num{50000} vehicles, the quality of rides and modal share improve a lot.
Fleets larger than \num{50000} vehicles provide only a small benefit to the quality of rides and the modal share compared to smaller fleets.
For example, the modal share of ride-pooling is \num{41.1}\% at \num{50000} vehicles of capacity four (with meeting points) and only improves to \num{45.5}\% for \num{500000} vehicles.

For small vehicle fleets, shared rides are vital, and we see high occupancy rates of up to \num{1.57}, \num{2.72}, and \num{3.84} and an average number of co-riders of up to \num{0.82}, \num{2.25}, and \num{4.28} for capacity two, four, and eight, respectively.
The level of pooling drops with growing vehicle fleets, as more vehicles are available to serve riders individually, like a taxi.
However, even for the largest fleets, the occupancy rate remains greater than \num{1.67} for a capacity of four or eight.

As expected, less pooling leads to worse system effectiveness for large fleets.
Intriguingly, for capacity two, the system effectiveness initially improves with growing fleet size even though the level of pooling falls slightly.
We believe that this is due to ineffective pooling, where new rides are found that have viable detours, but do not actually group riders with sufficiently similar itineraries. 
In this case, the vehicle and existing passengers experience long detours, which leads to long vehicle operation times, in-vehicle times, and wait times.
For larger fleets, it is more likely that non-shared rides are available and pooling is only applied if there is a good match.
Thus, for a capacity of two, the best system effectiveness is found at \num{100000} vehicles, as vehicles are not forced to make ineffective detours, but some pooling is still required. 
In the future, we would like to work on a re-assignment procedure that could help group riders better for small fleets compared to the current greedy approach.
\begin{figure}[tb]
\FIGURE
{
    \input{fig/tikzplots/Karlsruhe_full_day_paxdist}
}{
    Vehicle occupancy over time.
}{
    Shows distribution of vehicles over number of passengers and time for a full day variant of \KA with \num{25000} vehicles of capacity \num{4} (top), \num{25000} vehicles of capacity \num{8} (middle), and \num{50000} vehicles of capacity \num{4} (bottom).
    The height of each color stripe at a given point on the x-axis represents the number of vehicles that are currently idle (\textit{idle}) or driving with $k$ passengers (\textit{$k$ pax}).
}{
  \label{fig:KA_day_paxdist_plot}
}
\end{figure}

In~\Cref{fig:KA_day_paxdist_plot}, we show the occupancy of vehicles over time for a variant of the \KA instance that entails an entire day (from Tue 3 am to Wed 3 am).
We observe that the number of vehicles driving with three or four passengers follows the level of demand with a morning and afternoon peak.
When there is less demand in the middle of the day and in the evening, there is less stress on the system and more trips with one or two passengers can be made.

\subsubsection{Comparison between Urban Core and Periphery}
\label{subsubsec:core_and_periphery}
\begin{figure}[t]
\FIGURE
{
    \input{fig/tikzplots/KA_periphery_modes_plot}
}{
    Modal split classified by urban/rural origin and destination.
}{
    Fraction of travelers per mode of transportation (PT = public transit, RP = ride-pooling) for the \KA instance without ride-pooling (left) and with ride-pooling (right, \num{50000} vehicles).
    Shows distribution across modes for all requests (all), and subsets of urban-to-urban (u-u), rural-to-rural (r-r), urban-to-rural (u-r), and rural-to-urban (r-u) requests according to the RegioStaR-2 classification of Germany (see~\Cref{subsec:definition_core_periphery}).
}{
    \label{fig:distribution_modes_core_and_periphery}
}
\end{figure}

\begin{table}[t]
\TABLE
{
    Quality of trips classified by urban/rural origin and destination.
}{
    \begin{tabular}{@{}
    c
    l
    S[table-format = 7, table-alignment-mode = format, table-number-alignment=center]
    r
    r
    r
    r
    r
    S[table-format = 2.2, table-alignment-mode = format, table-number-alignment=center]
    @{}}
    \toprule
    inst. & \makecell[c]{type} & \multicolumn{1}{c}{\makecell[c]{\# req.}} & \makecell[c]{$\delta$\\{}\scriptsize[mm:ss]} & \makecell[c]{$\delta_{\textit{RP}}$\\{}\scriptsize[mm:ss]} & \makecell[c]{wait\\{}\scriptsize[mm:ss]} & \makecell[c]{ride (/$\delta_{\text{RP}}$)\\{}\scriptsize[mm:ss]} & \makecell[c]{trip\\{}\scriptsize[mm:ss]} & \multicolumn{1}{c}{\makecell[c]{RP\\{}\scriptsize[\%]}} \\
    \midrule
    \multirow{5}{*}{\KA}  & all & 975772 & 11:56 & 13:29 & 04:00 & 17:06 (1.20) & 22:11 & 41.09 \\
    & urban-urban & 703190 & 11:12 & 12:44 & 03:57 & 16:07 (1.20) & 21:10 & 40.14 \\
    & rural-rural & 184507 & 09:21 & 10:49 & 03:44 & 13:30 (1.18) & 18:14 & 40.64 \\
    & urban-rural & 36562 & 21:09 & 21:13 & 04:56 & 27:26 (1.27) & 33:29 & 49.43 \\
    & rural-urban & 51513 & 24:37 & 24:10 & 04:36 & 31:15 (1.27) & 36:55 & 49.73 \\
    \midrule
    \multirow{5}{*}{\ST} & all & 1388867 & 16:11 & 18:51 & 11:44 & 26:36 (1.35) & 39:58 & 32.52 \\
    & urban-urban & 1096587 & 15:14 & 17:32 & 11:24 & 24:48 (1.34) & 37:52 & 31.22 \\
    & rural-rural & 171481 & 12:20 & 15:19 & 11:28 & 20:53 (1.31) & 34:01 & 32.81 \\
    & urban-rural & 39519 & 33:03 & 34:05 & 14:21 & 48:12 (1.43) & 64:03 & 45.17 \\
    & rural-urban & 81280 & 29:01 & 29:34 & 13:57 & 42:20 (1.43) & 57:46 & 43.25 \\
    \bottomrule
    \end{tabular}
}{
    Key rider quality metrics for \mtkarri on the \KA and \ST instances using $|F| = \num{50000}$ vehicles.
    Rows show results for subsets of requests classified by origin and destination locale according to the RegioStaR-2 classification of Germany's low-level subdivisions into urban and rural areas (see~\Cref{subsec:definition_core_periphery}).
    Shows number of requests for each category, average car travel time from origin to destination of all requests ($\delta$) and only ride-pooling riders ($\delta_\text{RP}$), average wait time (wait), in-vehicle time (ride), and trip time (trip) of ride-pooling rides, as well as the percentage of travelers who choose ride-pooling (RP).
}{
    \label{tab:quality_core_and_periphery}
}
\end{table}

In this section, we evaluate the quality of ride-pooling for trips that move between the core and the periphery of a city.
This use case is particularly interesting for ride-pooling because peripheral areas provide fewer opportunities to use traditional public transit compared to urban areas.
However, urban areas attract a significant number of journeys from surrounding areas, e.g. for commuting to work or school.
Thus, the improved flexibility offered by ride-pooling may be a good opportunity to reduce the dominant use of private cars for these trips, and bundle the travelers into fewer vehicles for more efficient journeys.

We conduct experiments on the \KA and \ST instances with \num{50000} vehicles. 
For every origin and destination location in the request set, we decide whether it is urban or rural according to the RegioStaR-2 classification of Germany (for details, see~\Cref{subsec:definition_core_periphery}).
Requests fall into one of four categories: urban-urban, rural-rural, urban-rural, and rural-urban.
We run the dispatcher for all requests and then analyze the results by category, so requests of different categories affect each other's results like in a real system.
We report average results for three runs of each experiment.

In~\Cref{fig:distribution_modes_core_and_periphery}, we show the distribution of travelers across modes for all requests of \KA and for each of the four categories of requests.
The same plot for \ST can be found in~\Cref{fig:ST_distribution_modes_core_and_periphery} in~\Cref{sec:omitted_plots}.

The plot on the left shows the mode split when ride-pooling is not available.
For the urban-urban and rural-rural categories, many travelers choose to walk since travel distances are small.
In the urban-urban setting, about \num{18}\% of travelers choose public transit.
In the rural-rural setting, this number is lower at about \num{13}\%.
The urban-rural and rural-urban categories contain journeys with much larger travel distances, so few travelers walk.
Here, the modal share of the car is up to \num{66}\% and public transport accounts for more than \num{28}\% of trips.

The plot on the right shows the mode split including ride-pooling.
We find that ride-pooling takes on the largest share of any mode in all categories.
Ride-pooling seems to replace public transport as a more convenient alternative to private cars, and the share of transit falls to a single digit percentage in all categories. 
Importantly, ride-pooling is the most attractive in the urban-rural and rural-urban categories where cars previously dominated due to a lack of good alternatives.
Here, the introduction of ride-pooling causes the mode share of the car to fall from \num{66}\% to less than \num{45}\%. 

To understand the reason for the improved attractiveness of ride-pooling in urban-to-rural and rural-to-urban requests, we provide data on the quality of ride-pooling rides by category in \Cref{tab:quality_core_and_periphery}.
Note that rides in all categories have similar wait times and in-vehicle times relative to the shortest-path distance.
This seems to imply a similar quality of rides across all categories.
We find that the increased modal share in the urban-rural and rural-urban categories can instead be explained by longer travel distances compared to other categories (column $\delta$).
In our model, the probability of choosing ride-pooling grows with increasing travel distance since the wait time loses significance relative to the in-vehicle time.
For long trips, travelers decide that the discomfort of having to wait for a ride-pooling vehicle is outweighed by the fact that they do not have to operate a car themselves.
This supports the idea that ride-pooling can reduce the number of inefficient private car journeys between the center and the periphery of an urban area.
In the future, a multi-modal combination of ride-pooling and public transit may optimally leverage both modes for these trips.

\myparagraph{Fairness}
\begin{figure}[t]
\FIGURE
{
    \input{fig/tikzplots/periphery_cumulative_wait_and_ride_distribution}
}{
    Distribution of wait times and ride detours by urban/rural origin and destination.
}{
    Top: Cumulative distribution of wait times and ride detours of performed rides on \KA with \num{50000} vehicles.
    Bottom: Cumulative distribution of wait time and ride detour added to the originally offered ride after the rider has accepted the ride.
    We omit the top 0.1\% of values to improve readability.
    Each line shows the distribution for one category of requests.
}{
  \label{fig:periphery_wait_and_ride_distribution}
}
\end{figure}
Finally, we analyze the fairness of ride quality across our categories.
In~\Cref{fig:periphery_wait_and_ride_distribution}, we show plots on the cumulative distribution of wait times and ride detours relative to the direct car travel time, as well as the wait time and ride detour added after the rider accepted the offer.

First, consider the distribution of wait times.
We find that wait times have a similar distribution for all categories with a slightly more pronounced tail for the urban-rural and rural-urban categories.
This can again be attributed to the longer travel distance of requests in these categories.
Here, the travel time dominates the utility of each mode in the mode choice.
Thus, more rides with longer wait times will be accepted than in the other categories where travel times tend to be shorter.

We see similar results when considering the ride detour.
Note that the ride detour is smaller than zero if the traveler walks part of the total distance on their way to a meeting point.
This effect is more pronounced for urban-urban and rural-rural journeys where the total distance is smaller.
In addition, a larger fraction of rides have a ride detour of zero (i.e., non-shared rides) in these categories.
Due to the longer total travel distance in the urban-rural and rural-urban categories, larger detours are allowed and the riders occupy the vehicle for a longer time, so shared rides become more likely.
This results in slightly increased ride detours compared to the other categories.

For the vast majority of rides in all categories, the wait time is not extended beyond the wait time of the original ride offer, because it is unlikely that a detour is made between the points in time when the ride is accepted and when the rider is picked up. 
However, it is slightly more common in the urban-rural and rural-urban categories due to the longer wait times of accepted ride offers. 

The added ride detour compared to the original ride offer reflects our insight on the total ride detour.
In the urban-urban and rural-rural categories, fewer riders experience additional detours than in the urban-rural and rural-urban categories.
In the latter categories, almost two thirds of rides experience additional detours that are not expected at the time of accepting the ride offer.
However, in about 90\% of all rides, the additional detour stays below half the direct car travel time.

In conclusion, we find that journeys in the urban-urban and rural-rural categories experience slightly better rides than those in the urban-rural and rural-urban categories, in particular due to fewer shared rides.
However, journeys within the urban environment do not monopolize the ride-pooling fleet to the detriment of journeys in the periphery as we see similar distributions of wait times and ride detours for all categories.
Our mode choice model ignores wait times and ride detours added after accepting a ride, but riders would come to expect the additional inconvenience in a production system.
Since these added wait and travel times are more pronounced for urban-rural and rural-urban trips, this effect should be considered in more detail when designing a ride-pooling system, e.g., with respect to the service area and the pricing scheme.

\section{Conclusion}
\label{sec:conclusion}
Our dynamic ride-pooling dispatcher \mtkarri scales to millions of traveler requests per hour and allows experiments at an unprecedented scale.
In a simulation study, we find that a larger customer base can improve system effectiveness but may lead to system-wide degradation of ride-quality if every customer is always offered a ride.
Our experiments suggest that the use of meeting points, i.e., riders walking small distances to a pickup/dropoff location, improves pooling and can reduce the stress on a vehicle fleet in situations with high demand.
Furthermore, we show that more and/or larger vehicles lead to improved ride quality at the cost of decreased system effectiveness, but that a careful choice of the vehicle fleet can provide near-optimal rides while retaining good effectiveness.
An analysis of ride-pooling in peripheral areas suggests that it can be an attractive alternative to private cars in areas with limited choices of transportation modes.
We consider \mtkarri to be well-suited for further, more detailed studies on ride-pooling at a large scale. 

\subsection{Possible Extensions of the Algorithm and Simulation}
Our dispatching algorithm can be extended in different ways.

First, ride-pooling may be useful for feeder trips for traditional line-based public transit.
This use case might lead to high pooling potential as many passengers will go to railway stations or important bus stops.
It would be interesting to design an algorithm that computes optimal multi-modal journeys that contain stretches of ride-pooling and traditional transit. 
For this purpose, we could integrate \mtkarri as a transfer method in the ULTRA public transit routing framework~\citep{Baum2023}.

Furthermore, the algorithm could be adapted to incorporate changing travel times in a road network, e.g., to model updates for traffic congestion.
\karri permits the use of \emph{customizable} contraction hierarchies~\citep{Dibbelt2016}, which allow updates of all edge travel times in a road network in fractions of a second.
However, a change in travel times should also cause vehicles to adapt their current routes and could prompt a re-assignment of already assigned riders.
A dispatcher that incorporates this ability may strengthen the system's robustness. 

There is already a variant of \karri called KaRRiT~\citep{Breitling2025} that allows the dispatcher to offer rides with transfers between ride-pooling vehicles. 
Although KaRRiT is much slower than \karri without transfers, we would like to evaluate its performance with an added mode choice model on larger input instances like in this paper.

Currently, \mtkarri considers any location within walking distance to be a viable pickup or dropoff location.
In reality, not every location is suitable due to, e.g. poor visibility, lack of curb space, dangerous traffic conditions, etc.
Instead, there should be a fixed set of eligible locations in a road network that are known to be suitable as meeting points and that cover the network well.
It is an open question how these meeting points can be identified as part of an automated process.  

We might achieve higher quality by processing batches of requests in a more informed way. 
For example, for each request we could compute several possible vehicles and then solve a minimum weight perfect matching problem to select vehicles.
More sophisticated models might handle the case of assigning multiple requests to a single vehicle.

Finally, a tight coupling of the mobiTopp simulation framework and \mtkarri would allow for a direct injection of travel times and fleet status and thus a qualitatively improved mobility simulation.

\appendix

\section{Details on Generating Input Data}
\label{sec:details_on_generating_input_data}
In this appendix, we provide details on how we transformed our input data to generate ride-pooling instances.

\subsection{Departure Times}
\label{subsec:departure_times}
Our dispatcher supports departure times at a resolution of one second.
At this resolution, the distribution of travel times in the demands is skewed, with peaks every five minutes for the Karlsruhe and Stuttgart demands and every 15 minutes for the Los Angeles demand.
For example, when we divide the Los Angeles demand into intervals of 15 minutes, more than half of all travel requests in an average interval occur in the first second of the interval.
Presumably, this is an artifact of the transport simulations and the underlying activity plans, which are not meant to be used second-by-second.

Therefore, we generate a set of new request times that prevents these unnatural spikes but adheres to the original distribution at a resolution of 15-minute intervals as follows.
First, we divide the observation period into 15-minute intervals, shifting the start of each interval by 7.5 minutes so that each spike is in the middle of an interval.
For each interval, we count the number of requests.
Let $n_i$ be the number of requests in interval $i$.
The mean number of requests per second in batch $i$ is $m_i = n_i / 900$.
We assign the mean as a weight to the middle second $900i$ of each batch $i$ and interpolate linearly between these values.
More precisely, for second $x = 900i + c$ with $0 \le c < 900$, we assign weight $w(x) = m_i + \frac{m_{i+1} - m_i}{900} \cdot c$.
The sum of weights $\sum_x w(x)$ is equal to the number of requests.
Thus, we could simply assign $w(x)$ requests to second $x$.
However, this would be an unrealistically smooth distribution. 
Therefore, we randomly sample new request times from a cumulative distribution for every second $x$ of the observation period with probabilities $p(x) = x / \sum_{x'} w(x')$.
We sort the sampled request times and assign them to the requests in the original order.
While the expected number of requests per second $x$ is still $w(x)$, this randomization adds some nuance to the distribution and the resulting batches for the \mtkarri dispatcher.

\subsection{Generating Road Networks and Determining Possible Meeting Points}
\label{subsec:generating_road_networks_and_determining_meeting_points}
In this section, we explain how we construct the vehicle road network $\Gveh = (\Vveh, \Eveh)$ and the pedestrian road network $\Gpsg=(\Vpsg, \Epsg)$ in practice.
We give an overview over the vehicle networks used in~\Cref{tab:networks}.
We also describe how we map the locations of each travel request onto the generated networks.

\myparagraph{Generating Vehicle and Pedestrian Networks}
OpenStreetMap (OSM) (\url{https://openstreetmap.org}) provides detailed map data for many densely populated regions of the world, particularly in Western industrialized countries, including Germany and the United States. 
We base our road networks on publicly accessible OSM data obtained from \url{https://download.geofabrik.de}.
Next to regular roads, the OSM data includes footpaths, cycle paths, bridleways, etc., which we can use to determine a network of paths that are accessible to pedestrians.
For this, we rely on a classification of roads and paths into different categories used by OSM.

Every recorded road and path is called a \emph{highway} in OSM.
Every highway is attributed with a type that gives a high-level classification of its properties such as \texttt{motorway} or \texttt{residential} but also \texttt{footway} or even \texttt{steps}.
We ignore highways of more exotic types like \texttt{escape} or \texttt{via\_ferrata}.
The common classifications provide a reasonable indication of vehicle and pedestrian accessibility.
For the vehicle network $\Gveh$, we include any OSM highways (links) of the following types [\texttt{motorway}, \texttt{trunk}, \texttt{primary}, \texttt{secondary}, \texttt{tertiary}, \texttt{unclassified} (this type does \emph{not} indicate missing information), \texttt{residential}, and \texttt{living\_street}].
We exclude \texttt{service}, which describes access roads that a transit vehicle should not enter.
For the pedestrian network $\Gpsg$, we include any OSM highways of the following types [\texttt{tertiary}, \texttt{unclassified}, \texttt{residential}, \texttt{living\_street}, \texttt{service}, \texttt{pedestrian}, \texttt{track}, \texttt{footway}, \texttt{bridleway}, \texttt{steps}, \texttt{path}, and \texttt{cycleway}].
We include \texttt{bridleway} and \texttt{cycleway} because multi-use paths for horses/bicycles and pedestrians are often tagged as such.
Links of the types: [\texttt{tertiary}, \texttt{living\_street}] are present in both $\Gveh$ and $\Gpsg$, which can be used as meeting points for vehicles and passengers.
We tried to restrict this further to only roads that are tagged as having sidewalks, but we found this information to be incomplete in a lot of places.

In accordance with the MATSim Open Los Angeles scenario, we extend every vehicle network to include important roads (OSM categories \texttt{tertiary} and higher) in an extended area around the area of operation, so that vehicles can use these roads for realistic shortest paths.
For the \texttt{LA-$p$\%} instances, we add all important roads in Southern California (parts of California south of about 35°45'N, following \url{https://download.geofabrik.de/north-america/us/california/socal.html}).
For \ST and \KA, we add all important roads within a $200\si{km} \times 200\si{km}$ box that is aligned with the coordinate grid and concentric with the network.
This includes roads in France.

Our algorithm needs to know a travel time $\ell(e)$ for every edge $e \in \Eveh$.
We obtain these travel times based on the length of the road segment and the speed limit encoded in the OSM map data.
If no speed limit is specified in OSM, we use reasonable default values based on the highway type.
For the pedestrian graph $\Gpsg$, we encode the length of each path segment instead of the travel time since each rider can specify their own walking speed. 

\myparagraph{Mapping Rider Requests onto Generated Networks}
The demand data (see~\Cref{subsec:ride_pooling_scenarios}) is not based on the same representation of the road network of the respective area as the networks generated by us.
The MATSim Open Los Angeles scenario provides the OSM data used for their simulation.
However, we have to derive a vehicle network and a pedestrian network with a mapping between the two networks, which is why we use input data from OSM directly to ensure that all necessary information is contained in the OSM source. 
The Karlsruhe and Stuttgart instances are based on proprietary networks.
Thus, for every demand set, we need to map the O-D pairs from the source network into the networks in our representation.
As a preprocessing step, we perform the following mapping once for every demand and write the request set for the output network to disk.

We call the network on which a demand is based the input network and we call our representation the output network. 
To map the O-D pairs, we use the coordinate information stored for each vertex in the input and output networks.
We construct a k-d tree~\citep{bentley1975multidimensional}, a nearest-neighbor index that contains the coordinates of all vertices in the output network as points in two-dimensional Euclidean space.
Given a traveler's origin or destination location $v$ in the input network and the associated coordinates, we query the k-d tree to find the closest vertex to $v$ in the output network.
K-d trees do not represent geodesic distances fully accurately in large networks.
For very large networks, we could use a vantage-point tree~\citep{yianilos1993data} instead of the k-d tree.

We want to ensure that the origin and destination locations in the output network are accessible both by pedestrians and motor vehicles to guarantee that a rider's origin and destination are valid pickup/dropoff locations, and that every traveler can potentially walk and reach other meeting points. 
We restrict the k-d tree to only contain vertices that are incident to at least one edge that is accessible in the vehicle and the pedestrian road network.
Note that this may cause some skew in results depending on the request sets and the OSM data on which the output network is built. 

\myparagraph{Finding Possible Meeting Points for a Request}
For every incoming request $r=(\orig, \dest, \treq)$, we run a Dijkstra search in $\Gpsg$ rooted at $\orig$ to compute the possible pickup locations.
This search uses the length of the edges as its distance metric instead of the travel time.
We can stop the search as soon as the traveler-specific maximum walking distance $\lwalkmax(r)$ is reached.
Every incoming edge of every vertex settled by this search that is accessible in the vehicle network becomes a potential pickup location in $\Pickups(r)$.
We analogously run a reverse Dijkstra search rooted at $\dest$ to find the set of dropoff locations $\Dropoffs(r)$.
We use the walking speed $\walkspeed(r)$ to transform the distances computed by the Dijkstra searches into walking times for each pickup and dropoff location.

\section{Tuning of \karri Model Parameters}
\label{sec:tuning_of_karri_model_parameters}

In this appendix, we analyze the effect of the model parameters of \mtkarri on the solution quality.
We consider the metrics outlined in~\Cref{subsec:metrics}.
We run experiments where one parameter is varied, while all other parameters are fixed at their default value, which is specified at the end of the respective section.
We perform these experiments for the \KA and \LAten instances with a fleet of \num{25000} vehicles (see~\Cref{subsec:ride_pooling_scenarios}).

\subsection{Detour Cost Parameter}
\label{subsec:tuning_detour_cost}
\begin{figure}[tbh]
\FIGURE{
    \input{fig/tikzplots/combined_vehicle_cost_scale_quality_plot}
}{
    Solution quality for varying detour cost parameters.
}{
    Shows RP modal share, ride detour, wait time, occupancy, average number of co-riders, and system effectiveness for varying detour cost parameters $\detourweight \in \{0, 1, 2, 3\}$ on the \KA and \LAten instances with \num{25000} vehicles.
    For an explanation of the metrics, see~\Cref{subsec:metrics}.
    Note the different y-axis scales.
}{
  \label{fig:quality_detour_cost_parameter}
}
\end{figure}
The vehicle cost parameter $\detourweight$ determines the impact of the vehicle detour on the cost of an insertion.

We show how different values of $\detourweight$ affect our quality metrics in the plots in~\Cref{fig:quality_detour_cost_parameter}.
At $\detourweight = 0$, the cost of an insertion ignores the vehicle detour.
Thus, ride detours are small but routes become inefficient with little pooling and bad system effectiveness.
For \KA, the modal share of ride-pooling decreases with increasing $\detourweight$ as ride detours and wait times become longer.
For \LAten, ride detours also grow with larger $\detourweight$ but the wait time initially falls with larger $\detourweight$, so that the maximum modal share is reached at $\detourweight = 1$.
The decrease in wait times is caused by detours contributing to the wait time of currently waiting riders, so fewer and more efficient detours at larger $\detourweight$ lead to smaller average wait times. 
We do not see this effect in the \KA instance, as the level of pooling is much higher than for \LAten, and the longer wait times associated with shared rides compensate for the advantage of fewer and more efficient detours.

Increasing $\detourweight$ to values greater than one improves system effectiveness only slightly, but further decreases the attractiveness of ride-pooling.
Thus, we use $\detourweight = 1$ in our experiments to strike a balance between service quality and vehicle resource usage.

\subsection{Trip Time Constraint Parameters}
\label{subsec:tuning_trip_time_constraint_parameters}
\begin{figure}[tbh]
\FIGURE{
    \input{fig/tikzplots/combined_trip_time_constraint_quality_plot}
}{
    Solution quality for varying trip time constraint parameters.
}{
    Shows RP modal share, ride detour, wait time, occupancy, average number of co-riders, and system effectiveness for varying trip time constraint parameters $\alpha \in \{1.1, 1.2, 1.3, 1.4, 1.5\}$ and $\beta \in \{ 0, 5, 10, 15, 20 \}$ (in minutes) on the \KA and \LAten instances with \num{25000} vehicles.
    For an explanation of the metrics, see~\Cref{subsec:metrics}.
    Note the different y-axis scales.
}{
    \label{fig:quality_trip_time_constraint_parameters}
}
\end{figure}
The trip time constraint parameters $\alpha$ and $\beta$ define a threshold $\ttripmax(r) = \alpha \cdot T_t + \beta$ for the total trip time of a traveler $r$ relative to the tentative trip time $T_t$ of the ride that rider $r$ accepted.
If an insertion causes the trip time of any existing rider $r$ to exceed $\ttripmax(r)$, the insertion is considered infeasible.
Thus, the trip time constraint limits the detour that is allowed for existing riders.

We show quality results for varying $\alpha$ and $\beta$ in~\Cref{fig:quality_trip_time_constraint_parameters}.
Larger values of $\alpha$ and $\beta$ define a less restrictive trip time constraint and allow more insertions to vehicles with existing passengers.
This is reflected in an increased modal share of ride-pooling and slightly longer ride detours, as well as shorter wait times due to a larger supply of possible rides in the vicinity.
The level of pooling increases with a greater average occupancy and number of co-riders since detours for shared rides are more likely to be feasible.
The increase in pooling causes more efficient use of vehicles and an improved system effectiveness.

With sufficiently large $\alpha$ and/or $\beta$, each metric approaches an asymptote.
We choose to use $\alpha = 1.4$ and $\beta = 10\si{min}$, which is close to this asymptote, to prioritize system-wide quality expressed by a large modal share and good system effectiveness.
Note that this differs from not using a trip time constraint at all since the constraint maintains a guaranteed service quality for every individual rider.

\subsection{Maximum Waiting Time Parameter}
\label{subsec:tuning_maximum_wait_time}
\begin{figure}[tbh]
\FIGURE{
    \input{fig/tikzplots/combined_max_wait_time_quality_plot}
}{
    Solution quality for varying maximum wait time.
}{
    Shows RP modal share, ride detour, wait time, occupancy, average number of co-riders, and system effectiveness for varying maximum wait time $\twaitmax \in \{0, 2.5, 5, 7.5, 10, 12.5\}$ (in minutes) on the \KA and \LAten instances with \num{25000} vehicles.
    For an explanation of the metrics, see~\Cref{subsec:metrics}.
    Note the different y-axis scales.
}{
    \label{fig:quality_max_wait_time}
}
\end{figure}
The maximum waiting time $\twaitmax$ comprises a threshold for the waiting time of a rider, i.e., the time between issuing the request and being picked up by a vehicle.
Assume that an existing rider $r$ accepted a ride with a tentative pickup time of $T_p$.
If an insertion causes the pickup time of $r$ to exceed $T_p + \twaitmax$, the insertion is considered infeasible.

We show the effect of $\twaitmax$ on the solution quality of \mtkarri in~\Cref{fig:quality_max_wait_time}.
The maximum wait time parameter has only a small effect on every metric.
This is likely due to the wait time constraint only having an effect in the rare case that an insertion causes a detour on the way to picking up an already assigned rider.
A value of $\twaitmax = 0$ does not allow for this case at all.
This makes too many insertions infeasible, and we actually see longer wait times and a slightly smaller modal share than for larger values of $\twaitmax$.

We choose $\twaitmax = 10\si{min}$ to maximize modal share.

\subsection{Walking Cost Parameter}
\label{subsec:tuning_walking_cost}
\begin{figure}[tbh]
\FIGURE{
    \input{fig/tikzplots/Karlsruhe_walking_cost_scale_quality_plot}
}{
    Solution quality for varying walking cost parameters.
}{
    Shows RP modal share, ride detour, wait time, occupancy, average number of co-riders, and system effectiveness for varying walking cost parameters $\walkweight \in \{0, 1, 2, 3\}$ on the \KA instance with \num{25000} vehicles.
    To determine meeting points, we use the rider-specific walking speed and maximum walking distance given by the \KA instance.
    For an explanation of the metrics, see~\Cref{subsec:metrics}.
    Note the different y-axis scales.
}{
  \label{fig:quality_walking_cost_parameter}
}
\end{figure}
The walking cost parameter $\walkweight$ determines the impact of walking on the cost of an insertion.
The walking time includes the time the rider spends walking from their origin to the pickup location of the insertion and from the dropoff location of the insertion to their destination.
With greater values of $\walkweight$, a short walking time becomes more important.

The plots in~\Cref{fig:quality_walking_cost_parameter} show how different quality metrics are affected by the walking cost parameter $\walkweight$.
The plot only considers the \KA instance, since we do not allow meeting points for \LAten.

The best RP modal share and average occupancy are achieved at $\walkweight = 1$.
This comes at a cost of increased ride detours and wait times, and slightly worse system effectiveness compared to $\walkweight=0$.
All metrics become worse for $\walkweight > 1$.
Note that the wait time does not include the walking time to the pickup location.

\begin{figure}[tbh]
\FIGURE{
    \input{fig/tikzplots/Karlsruhe_walking_times}
}{
    Walking times for varying walking cost parameters.
}{
    Shows absolute walking times (left) and normalized walking times (right) for varying walking cost parameters $\walkweight \in \{0, 1, 2, 3\}$ on the \KA instance with \num{25000} vehicles.
    The normalized walking time on the right is defined as the median of the ratios between the walking time and the shortest-path travel time from origin to destination by car for every rider.
}{
    \label{fig:walking_times_walking_cost_parameter}
}
\end{figure}
It is important to also consider the effect on the actual walking times.
The left plot in~\Cref{fig:walking_times_walking_cost_parameter} shows the average absolute walking times of riders for varying $\walkweight$.
The walking time falls with increasing $\walkweight$.
At $\walkweight=0$, each rider walks for $121\s$ on average, which drops to around $78\s$ at $\walkweight=1$.

The y-axis of the right plot in~\Cref{fig:walking_times_walking_cost_parameter} shows the median of the ratios between the walking time and the shortest-path travel time by car.
The plot shows that at $\walkweight=0$, in the median, the walking time of a rider is almost $16\%$ of the entire trip from origin to destination by car.
This ratio decreases to less than $10\%$ for $\walkweight = 1$.

To ensure the maximum RP modal share and limited walking times, we choose to use $\walkweight=1$ in our experiments.

\section{Strong Scaling}
\label{sec:strong_scaling}
\begin{table}[t]
\TABLE
{
    Strong scaling of \mtkarri.
}{
    \sisetup{}
    \begin{tabular}{@{}
    S[table-format = 2, table-alignment-mode = format, table-number-alignment=right] 
    S[table-format = 3, table-alignment-mode = format, table-number-alignment=center]
    S[table-format = 1, table-alignment-mode = format, table-number-alignment=center]
    S[table-format = 2, table-alignment-mode = format, table-number-alignment=center]
    S[table-format = 3, table-alignment-mode = format, table-number-alignment=center]
    S[table-format = 5, table-alignment-mode = format, table-number-alignment=center] 
    S[table-format = 1, table-alignment-mode = format, table-number-alignment=center]
    S[table-format = 3, table-alignment-mode = format, table-number-alignment=center]
    S[table-format = 5, table-alignment-mode = format, table-number-alignment=center]
    @{}}
    \toprule
         & \multicolumn{4}{c}{\LAone} & \multicolumn{4}{c}{\LAtwentyfive} \\
         \cmidrule(lr){2-5}\cmidrule(l){6-9}
         \multicolumn{1}{c}{$T$} & \multicolumn{1}{c}{ins.} & \multicolumn{1}{c}{mode} & \multicolumn{1}{c}{update} & \multicolumn{1}{c}{total} & \multicolumn{1}{c}{ins.} & \multicolumn{1}{c}{mode} & \multicolumn{1}{c}{update} & \multicolumn{1}{c}{total} \\
         \cmidrule(r){1-1}\cmidrule(lr){2-5}\cmidrule(l){6-9}
         \multicolumn{1}{c}{no batches} &
                553 & 3 & 89 & 646 & 8363 & 5 & 1123 & 9491 \\
         \cmidrule(r){1-1}\cmidrule(lr){2-5}\cmidrule(l){6-9}
            1 & 576 & 7 & 97 & 680 & 9460 & 7 & 256 & 9724 \\
            4 & 153 & 8 & 48 & 208 & 2206 & 7 &  93 & 2306 \\
            8 &  88 & 7 & 41 & 137 & 1253 & 7 &  68 & 1328 \\
           16 &  61 & 8 & 53 & 122 &  643 & 7 &  79 &  729 \\
           32 &  51 & 8 & 55 & 114 &  336 & 7 &  76 &  420 \\
           64 &  54 & 8 & 60 & 121 &  203 & 7 &  76 &  286 \\
           96 &  60 & 8 & 65 & 133 &  165 & 7 &  77 &  250 \\
    \bottomrule
    \end{tabular}
}{
    Average running times per request (in $\mus$) for \karri without batching (\emph{no batches}) and \mtkarri with different numbers of threads ($T$) on \LAone (left) and \LAtwentyfive (right) with \num{5000} and \num{50000} vehicles, respectively.
    Shows average running time per request for finding the best ride-pooling insertion (ins.), choosing the mode of transportation (mode), updating the system state of the ride-pooling dispatcher (update), and the total time (total).
}{
    \label{tab:absolute_running_times_strong_scaling}
}
\end{table}

\begin{figure}
\FIGURE
{
    \input{fig/tikzplots/LA-1-and-25-pct_speedups_plot}
}{
    Speedups of \mtkarri over the non-batched sequential version.
}{
    Speedups of \mtkarri over sequential \karri with different number of threads (logarithmic x-axis) on \LAone with \num{5000} vehicles (left) and \LAtwentyfive with \num{50000} vehicles (right).
    Shows separate speedups for finding insertions, updating the system state, and the total running time.
    We do not show speedups for mode choice, since we did not parallelize it.
    Note the different y-axis scales.
}{
    \label{fig:speedups}
}
\end{figure}
\begin{figure}
\FIGURE
{
    \input{fig/tikzplots/LA-1-and-25-pct_component_fraction_plot}
}{
    Composition of running times by component for different numbers of threads.
}{
    Fraction of running time per component of non-batched \karri (\emph{no b.}) and \mtkarri with different numbers of threads $T$ on \LAone (left) and \LAtwentyfive (right) with \num{5000} and \num{50000} vehicles, respectively.
}{
    \label{fig:component_fraction_of_running_time}
}
\end{figure}

We analyze the performance of \mtkarri for an increasing number of threads on \LAone and \LAtwentyfive with \num{5000} and \num{50000} vehicles, respectively.
We run \mtkarri using $T \in \{1,4,8,16,32,64,96\}$ threads and evaluate the speedups compared to sequential \karri.
We perform only a single run per configuration since the runs with smaller numbers of threads take a long time.

We show absolute running times in~\Cref{tab:absolute_running_times_strong_scaling} and speedup plots in~\Cref{fig:speedups}.
We did not parallelize the mode choice at all, so we do not show this component in the plots.
We only parallelized some parts of the system state updates, which is why there are no or very limited speedups for this component.
We focused our efforts on the computation of insertions.
For the smaller \LAone instance, there is only an average of \num{39.7} requests in every batch, which is not enough to utilize all threads or provide good load balancing (we use the default of $\tbatch=5\s$ here).
Thus, the speedups for the computation of insertions is limited to a factor of at most \num{10.80} for \LAone, which is reached at $T=32$ threads
The larger instance provides larger batches with an average of \num{981.6} requests per batch.
With this, \mtkarri with $T=96$ threads computes insertions \num{50.54} times faster than \karri.
The total speedup is at most \num{5.66} and \num{38.01} at $T=32$ and $T=96$ threads on the \LAone and \LAtwentyfive instances, respectively.

As shown in~\Cref{fig:component_fraction_of_running_time}, the mode choice and updates of the system state make up only about 14.3\% and 11.8\% of the total running time in the sequential setting for the smaller and larger instances, respectively.
This is why we initially focused our parallelization efforts on the dominant task of finding insertions for every request.
However, with $T=\num{96}$ threads the mode choice and the system state update contribute a significant part of the running time.
In the future, we aim to improve the parallelization of these components to allow for good scalability to even larger numbers of threads.

\section{Omitted Plots}
\label{sec:omitted_plots}
In this appendix, we include plots for the \ST instance that were omitted from the main part of the paper due to space limitations.
\Cref{fig:fleet_quality_plot_stuttgart} shows a plot on the solution quality for different fleet sizes.
It is analogous to~\Cref{fig:fleet_quality_plot_karlsruhe} in~\Cref{subsubsec:impact_of_fleet}.
\Cref{fig:ST_distribution_modes_core_and_periphery} shows a plot on the modal split with and without ride-pooling.
It is analogous to~\Cref{fig:distribution_modes_core_and_periphery} in~\Cref{subsubsec:core_and_periphery}.
\begin{figure}[tb]
\FIGURE
{
    \input{fig/tikzplots/Stuttgart_fleet_quality_plot}
}{
    Effect of fleet size on quality.
}{
    Quality on \ST with fleets $F$ of $|F| \in \{ \num{10}, \num{20}, \num{30},\num{40},\num{50},\num{75},\num{100}, \num{150}, \num{200}, \num{300}, \num{400}, \num{500} \} \cdot 10^3$ vehicles and $\capacity(\nu) \in \{2,4,8\}$ for every $\nu \in F$ with and without meeting points (\textit{MPs}). 
    The x-axis describes the number of vehicles in thousands. 
    The y-axis shows (left to right, top to bottom) RP modal share, ride detour, average wait time, occupancy, number of co-riders, and system effectiveness, respectively.
    For explanations of the metrics, see~\Cref{subsec:metrics}.
    Note the different y-axis scales.
}{
    \label{fig:fleet_quality_plot_stuttgart}
}
\end{figure}
\begin{figure}[t]
\FIGURE
{
    \input{fig/tikzplots/ST_periphery_modes_plot}
}{
    Modal split for \ST classified by urban/rural origin and destination.
}{
    Fraction of travelers per mode of transportation (PT = public transit, RP = ride-pooling) for the \ST instance without ride-pooling (left) and with ride-pooling (right, \num{50000} vehicles).
    Shows distribution across modes for all requests (all), and subsets of urban-to-urban (u-u), rural-to-rural (r-r), urban-to-rural (u-r), and rural-to-urban (r-u) requests according to the RegioStaR-2 classification of Germany (see~\Cref{subsec:definition_core_periphery}).
}{
  \label{fig:ST_distribution_modes_core_and_periphery}
}
\end{figure}
\bibliography{references}

@Article{Dijkstra1959,
  author	= {Edsger W Dijkstra},
  journal	= {Numerische Mathematik},
  title	= {A Note on Two Problems in Connexion with Graphs},
  volume	= {1},
  year	= {1959}
}

@Article{Geisberger2012,
  author	      = {Robert Geisberger and Peter Sanders and Dominik Schultes and Christian Vetter},
  doi		      = {10.1287/trsc.1110.0401},
  issue		  = {3},
  journal	      = {INFORMS Transportation Science},
  publisher	  = {INFORMS},
  title		  = {Exact Routing in Large Road Networks using Contraction Hierarchies},
  volume	      = {46},
  year		  = {2012}
}

@InProceedings{Knopp2007,
  author	= {Sebastian Knopp and Peter Sanders and Dominik Schultes and Frank Schulz and Dorothea Wagner},
  doi		= {10.1137/1.9781611972870.4},
  booktitle	= {Workshop on Algorithm Engineering and Experiments (ALENEX)},
  title		= {Computing Many-to-Many Shortest Paths using Highway Hierarchies},
  publisher = {SIAM},
  year		= {2007}
}

@Article{	  Dibbelt2016,
  author	= {Julian Dibbelt and Ben Strasser and Dorothea Wagner},
  doi		= {10.1145/2886843},
  issn		= {1084-6654},
  journal	= {ACM Journal of Experimental Algorithmics},
  pages		= {1-49},
  publisher	= {ACM},
  title		= {Customizable Contraction Hierarchies},
  volume	= {21},
  year		= {2016}
}

@Article{Baum2023,
    author = {Moritz Baum and Valentin Buchhold and Jonas Sauer and Dorothea Wagner and Tobias Zündorf},
    doi = {10.1287/trsc.2022.0198},
    journal = {INFORMS Transportation Science},
    publisher = {INFORMS},
    volume = {57},
    issue = {6},
    title = {{ULTRA}: Unlimited Transfers for Efficient Multimodal Journey Planning},
    year = {2023}
}

@InProceedings{Buchhold2021,
  author	     = {Valentin Buchhold and Peter Sanders and Dorothea Wagner},
  city	     = {Philadelphia, PA},
  doi		     = {10.1137/1.9781611976472.8},
  isbn	     = {9781611976472},
  issn	     = {21640300},
  booktitle	 = {Workshop on Algorithm Engineering and Experiments (ALENEX)},
  pages	     = {98-112},
  publisher    = {SIAM},
  title	     = {Fast, Exact and Scalable Dynamic Ridesharing},
  year	     = {2021}
}

@InProceedings{Laupichler2024,
author = {Moritz Laupichler and Peter Sanders},
title = {Fast Many-to-Many Routing for Dynamic Taxi Sharing with Meeting Points},
booktitle = {Symposium on Algorithm Engineering and Experiments (ALENEX)},
chapter = {},
pages = {74-90},
publisher={SIAM},
year={2024},
doi = {10.1137/1.9781611977929.6}
}

@article{bentley1975multidimensional,
author = {Bentley, Jon Louis},
title = {Multidimensional binary search trees used for associative searching},
year = {1975},
issue_date = {Sept. 1975},
volume = {18},
number = {9},
issn = {0001-0782},
url = {https://doi.org/10.1145/361002.361007},
doi = {10.1145/361002.361007},
journal = {Communications of the ACM},
month = sep,
pages = {509–517},
publisher = {ACM},
numpages = {9}
}

@inproceedings{yianilos1993data,
author = {Yianilos, Peter N.},
title = {Data structures and algorithms for nearest neighbor search in general metric spaces},
year = {1993},
isbn = {0898713137},
publisher = {SIAM},
booktitle = {Symposium on Discrete Algorithms (SODA)},
pages = {311-321},
numpages = {11},
location = {Austin, Texas, USA}
}

@article{bohannon1997comfortable,
  title={Comfortable and maximum walking speed of adults aged 20—79 years: reference values and determinants},
  author={Bohannon, Richard W},
  journal={Age and Ageing},
  volume={26},
  number={1},
  pages={15--19},
  year={1997},
  publisher={Oxford University Press}
}

@techreport{nobis2018MiD,
  title={{Mobilit{\"a}t in Deutschland- MiD: Ergebnisbericht}},
  author={Nobis, Claudia and Kuhnimhof, Tobias},
  year={2018},
  institution={{Bundesministerium für Verkehr und digitale Infrastruktur (BMVI)}}
}

@Book{		  Train2003,
  editor	= {Kenneth E. Train},
  isbn		= {0521816963},
  pages		= {334},
  publisher	= {Cambridge University Press},
  title		= {Discrete Choice Methods with Simulation},
  year		= {2003}
}

@techreport{BMDV2018RegioStaR,
  author       = {{Federal Ministry of Transport and Digital Infrastructure (BMVI)}},
  title        = {{RegioStaR} -- Regional Statistical Area Typology for Mobility and Transport Research},
  institution  = {Federal Ministry of Transport and Digital Infrastructure (BMVI)},
  year         = {2018},
  address      = {Berlin, Germany},
  url          = {https://www.bmdv.bund.de/SharedDocs/DE/Anlage/G/regiostar-raumtypologie-englisch.pdf},
  note         = {Accessed: 2025-11-28}
}

@techreport{BBSR2018_OeV_Angebotsqualitaet,
  author       = {{Bundesinstitut für Bau-, Stadt- und Raumforschung (BBSR)}},
  title        = {{Verkehrsbild Deutschland 2018: Die Angebotsqualität im Öffentlichen Verkehr (ÖV)}},
  institution  = {{Bundesinstitut für Bau-, Stadt- und Raumforschung (BBSR)}},
  year         = {2018},
  address      = {Bonn, Germany},
  type         = {Analysen Kompakt 08/2018},
  url          = {https://www.bbsr.bund.de/BBSR/DE/veroeffentlichungen/analysen-kompakt/2018/ak-08-2018-dl.pdf},
  note         = {Accessed: 2025-11-28}
}

@techreport{Belz2025,
           month = {November},
            year = {2025},
           title = {{M}obilit{\"a}t in {D}eutschland {--} {MiD} {E}rgebnisbericht},
          author = {Belz, Janina and B{\"o}hnen, Carina and Brand, Thorsten and Dubernet, Ilka and Follmer, Robert and Galich, Anton and Gruschwitz, Dana and H{\"o}ing, Niklas and J{\"a}kel, Klaus and Jearwattanakanok, Krongkwan and K{\"o}hler, Katja and Kuhnimhof, Tobias and van Nek, Lea and Nobis, Claudia and Plener, Martin and Porschen, Marcel and Ruppenthal, Merle and Schelewsky, Marc and Schuppan, Julia and Sch{\"u}rig, Michaela and Seitz, Ingmar and Wawrzyniak, Barbara},
          institution = {infas Institute for Applied Social Science and German Aerospace Center and IVT Research GmbH},
             doi = {10.71786/dlr.vf-9hvj-sh11},
             url = {https://elib.dlr.de/221542/},
        note = {commissioned by the Federal Ministry for Digital and Transport}
}

@techreport{Eurostat2018,
    author = {{Eurostat Task Force on Passenger Mobility}},
    title = {Passenger Mobility Statistics - Report on Surveys},
    institution = {Eurostat},
    year = {2018},
    url = {https://circabc.europa.eu/ui/group/097ec987-3401-48d2-8cff-925d158b6eb1/library/25861439-946c-4932-a8af-502d3eee7243/details},
    note = {Accessed: 2026-02-11}
}

@InProceedings{Breitling2025,
  author =	{Breitling, Johannes and Laupichler, Moritz},
  title =	{Exact and Heuristic Dynamic Taxi Sharing with Transfers Using Shortest-Path Speedup Techniques},
  booktitle =	{Symposium on Algorithmic Approaches for Transportation Modelling, Optimization, and Systems (ATMOS)},
  pages =	{15:1--15:22},
  ISBN =	{978-3-95977-404-8},
  ISSN =	{2190-6807},
  year =	{2025},
  publisher =	{Schloss Dagstuhl -- Leibniz-Zentrum f{\"u}r Informatik},
  URL =		{https://drops.dagstuhl.de/entities/document/10.4230/OASIcs.ATMOS.2025.15},
  URN =		{urn:nbn:de:0030-drops-247718},
  doi =		{10.4230/OASIcs.ATMOS.2025.15}
}

@techreport{ITF2015,
    author = {{International Transport Forum}},
    title = {Urban Mobility System Upgrade: {H}ow shared self-driving cars could change city traffic},
    institution = {OECD},
    year = {2015},
    doi = {10.1787/5jlwvzdk29g5-en}
}

@techreport{Fulton2017,
  title={Three revolutions in urban transportation: How to achieve the full potential of vehicle electrification, automation, and shared mobility in urban transportation systems around the world by 2050},
  author={Fulton, Lew and Mason, Jacob and Meroux, Dominique},
  institution={Institute for Transportation and Development Policy},
  year={2017},
  url = {https://steps.ucdavis.edu/wp-content/uploads/2017/05/ITDP-3R-Report-v6.pdf},
  note = {Accessed 2026-02-11}
}

@techreport{Garus2025,
	number = {KJ-01-25-260-EN-N},
	issn = {1831-9424},
	year = {2025},
	author = {A. Garus and K. Mattas and G. Albano and G. Baldini and B. Ciuffo},
    institution = {Joint Research Centre},
	isbn = {978-92-68-27059-2},
	publisher = {Publications Office of the European Union},
	title = {On the energy intensity of road transport in the presence of Connected and Automated Mobility},
	type = {},
	url = {},
	doi = {10.2760/8978379}
}

@techreport{UCDavis2024,
    title={Car Dependence in {Los Angeles}},
    author={{UC Davis and KIT}},
    institution={University of California, Davis and Karlsruhe Institute of Technology},
    year={2024},
    month={4},
    url = {https://3rev.ucdavis.edu/sites/g/files/dgvnsk14786/files/media/documents/LA_Survey_Exec_Summary_Final_Version.pdf},
    note = {Study supported by BMW group. Accessed 2026-02-25}
}

@Article{     Psaraftis1980,
  author	= {Harilaos N. Psaraftis},
  doi		= {10.1287/trsc.14.2.130},
  issn		= {0041-1655},
  issue		= {2},
  journal	= {Transportation Science},
  pages		= {130-154},
  title		= {A Dynamic Programming Solution to the Single Vehicle Many-to-Many Immediate Request Dial-a-Ride Problem},
  volume	= {14},
  year		= {1980}
}

@Article{	  Jaw1986,
  author	= {Jang-Jei Jaw and Amedeo R. Odoni and Harilaos N. Psaraftis
		  and Nigel H.M. Wilson},
  doi		= {10.1016/0191-2615(86)90020-2},
  issn		= {01912615},
  issue		= {3},
  journal	= {Transportation Research Part B: Methodological},
  pages		= {243-257},
  publisher	= {Elsevier},
  title		= {A Heuristic Algorithm for the Multi-Vehicle Advance
		  Request Dial-a-Ride Problem with Time Windows},
  volume	= {20},
  year		= {1986}
}

@Article{	  Madsen1995,
  author	= {Oli B. G. Madsen and Hans F. Ravn and Jens Moberg
		  Rygaard},
  doi		= {10.1007/BF02031946},
  issn		= {0254-5330},
  issue		= {1},
  journal	= {Annals of Operations Research},
  pages		= {193-208},
  publisher	= {Baltzer Science Publishers, Baarn/Kluwer Academic
		  Publishers},
  title		= {A Heuristic Algorithm for a Dial-a-Ride Problem with Time
		  Windows, Multiple Capacities, and Multiple Objectives},
  volume	= {60},
  year		= {1995}
}

@Article{	  Horn2002,
  author	= {Mark E.T. Horn},
  doi		= {10.1016/S0968-090X(01)00003-1},
  issn		= {0968090X},
  issue		= {1},
  journal	= {Transportation Research Part C: Emerging Technologies},
  pages		= {35-63},
  publisher	= {Elsevier},
  title		= {Fleet Scheduling and Dispatching for Demand-Responsive
		  Passenger Services},
  volume	= {10},
  year		= {2002}
}

@Article{	  Hunsaker2002,
  author	= {Brady Hunsaker and Martin Savelsbergh},
  doi		= {10.1016/S0167-6377(02)00120-7},
  issn		= {01676377},
  issue		= {3},
  journal	= {Operations Research Letters},
  pages		= {169-173},
  title		= {Efficient Feasibility Testing for Dial-a-Ride Problems},
  volume	= {30},
  year		= {2002}
}

@Article{	  Haell2012,
  author	= {Carl H. Häll and Magdalena Högberg and Jan T. Lundgren},
  doi		= {10.1007/s12469-012-0052-6},
  issn		= {1866-749X},
  issue		= {1},
  journal	= {Public Transport},
  pages		= {17-37},
  title		= {A Modeling System for Simulation of Dial-a-Ride Services},
  volume	= {4},
  year		= {2012}
}

@Article{	  Ota2017,
  author	= {Masayo Ota and Huy Vo and Claudio Silva and Juliana
		  Freire},
  doi		= {10.1109/TBDATA.2016.2627223},
  issn		= {2332-7790},
  issue		= {3},
  journal	= {IEEE Transactions on Big Data},
  pages		= {349-361},
  publisher	= {IEEE},
  title		= {{STaRS}: Simulating Taxi Ride Sharing at Scale},
  volume	= {3},
  year		= {2017}
}

@Article{	  Lotze2022,
  author	= {Charlotte Lotze and Philip Marszal and Malte Schröder and
		  Marc Timme},
  doi		= {10.1088/1367-2630/ac47c9},
  issn		= {1367-2630},
  issue		= {2},
  journal	= {New Journal of Physics},
  pages		= {023034},
  title		= {Dynamic stop pooling for flexible and sustainable ride
		  sharing},
  volume	= {24},
  year		= {2022}
}

@InProceedings{	  Bischoff2017,
  author	= {Joschka Bischoff and Michal Maciejewski and Kai Nagel},
  doi		= {10.1109/ITSC.2017.8317926},
  booktitle	= {International Conference on Intelligent Transportation Systems (ITSC)},
  title		= {City-Wide Shared Taxis: A Simulation Study in {Berlin}},
  publisher = {{IEEE}},
  year		= {2017}
}

@Book{		  Horni2016,
  doi		= {10.5334/baw},
  editor	= {Andreas Horni and Kai Nagel and Kay W. Axhausen},
  isbn		= {9781909188754},
  pages		= {618},
  publisher	= {Ubiquity Press},
  title		= {The Multi-Agent Transport Simulation {MATSim}},
  year		= {2016}
}

@article{Mallig2014,
title = {{mobiTopp} – A Modular Agent-based Travel Demand Modelling Framework},
journal = {Procedia Computer Science},
volume = {19},
pages = {854-859},
year = {2013},
issn = {1877-0509},
doi = {https://doi.org/10.1016/j.procs.2013.06.114},
url = {https://www.sciencedirect.com/science/article/pii/S1877050913007229},
author = {Nicolai Mallig and Martin Kagerbauer and Peter Vortisch}
}

@InProceedings{	  Huang2014,
  author	= {Yan Huang and Favyen Bastani and Ruoming Jin and Xiaoyang
		  Sean Wang},
  doi		= {10.14778/2733085.2733106},
  issn		= {21508097},
  issue		= {14},
  booktitle	= {Proceedings of the VLDB Endowment},
  pages		= {2017-2028},
  publisher	= {ACM},
  title		= {Large Scale Realtime Ridesharing with Service Guarantee on
		  Road Networks},
  volume	= {7},
  year		= {2014}
}

@inproceedings{Kleiner2011,
author = {Kleiner, Alexander and Nebel, Bernhard and Ziparo, Vittorio Amos},
title = {A mechanism for dynamic ride sharing based on parallel auctions},
year = {2011},
isbn = {9781577355137},
publisher = {AAAI},
booktitle = {International Joint Conference on Artificial Intelligence (IJCAI)},
pages = {266–272},
numpages = {7},
location = {Barcelona, Catalonia, Spain}
}

@article{Agatz2011,
   author = {Niels Agatz and Alan Erera and Martin Savelsbergh and Xing Wang},
   doi = {10.1016/j.trb.2011.05.017},
   issn = {01912615},
   issue = {9},
   journal = {Transportation Research Part B: Methodological},
   month = {11},
   pages = {1450-1464},
   publisher = {Elsevier},
   title = {Dynamic Ride-Sharing: A Simulation Study in Metro {Atlanta}},
   volume = {45},
   url = {https://linkinghub.elsevier.com/retrieve/pii/S0191261511000671},
   year = {2011}
}

@inproceedings{Herbawi2012,
   author = {Wesam Herbawi and Michael Weber},
   doi = {10.1145/2330163.2330219},
   isbn = {9781450311779},
   booktitle = {International Conference on Genetic and Evolutionary Computation (GECCO)},
   publisher = {ACM},
   pages = {385-392},
   title = {A Genetic and Insertion Heuristic Algorithm for Solving the Dynamic Ridematching Problem with Time Windows},
   year = {2012}
}

@techreport{worlen2025verkehrsentlastung,
  title={{Verkehrsentlastung durch neue Arbeitsformen und Mobilit{\"a}tstechnologien: Abschlussbericht VENAMO Projekt}},
  author={W{\"o}rlen, Matthias and Wist, Sarah and Reiffer, Anna and Hallensleben, Tobias and Kuhn, Rainer and Kagerbauer, Martin and Bofinger, Phillipp and Kandler, Kim},
  year={2025},
  institution={ETH Zurich}
}

@article{worle2021modeling,
  title={Modeling intermodal travel behavior in an agent-based travel demand model},
  author={W{\"o}rle, Tim and Briem, Lars and Heilig, Michael and Kagerbauer, Martin and Vortisch, Peter},
  journal={Procedia Computer Science},
  volume={184},
  pages={202--209},
  year={2021},
  publisher={Elsevier}
}

@InProceedings{	  Gilibert2020,
  author	= {Mireia Gilibert and Imma Ribas and Christian Rosen and
		  Alexander Siebeneich},
  doi		= {10.1016/j.trpro.2020.03.105},
  issn		= {23521465},
  booktitle	= {Transportation Research Procedia},
  pages		= {323-330},
  publisher	= {Elsevier},
  title		= {On-demand Shared Ride-hailing for Commuting Purposes:
		  Comparison of {Barcelona} and {Hannover} Case Studies},
  volume	= {47},
  year		= {2020}
}

@Article{	  Kostorz2021,
  author	= {Nadine Kostorz and Eva Fraedrich and Martin Kagerbauer},
  doi		= {10.3390/su13020958},
  issn		= {2071-1050},
  issue		= {2},
  journal	= {Sustainability},
  pages		= {958},
  publisher	= {MDPI},
  title		= {Usage and User Characteristics—Insights from {MOIA},
		  Europe’s Largest Ridepooling Service},
  volume	= {13},
  year		= {2021}
}

@Article{	  Yu2017,
  author	= {Biying Yu and Ye Ma and Meimei Xue and Baojun Tang and Bin
		  Wang and Jinyue Yan and Yi-Ming Wei},
  doi		= {10.1016/j.apenergy.2017.01.052},
  issn		= {03062619},
  journal	= {Applied Energy},
  pages		= {141-152},
  title		= {Environmental Benefits From Ridesharing: A Case of
		  {Beijing}},
  volume	= {191},
  year		= {2017}
}

@Article{	  Weckstroem2018,
  author	= {Christoffer Weckström and Miloš N. Mladenović and Waqar
		  Ullah and John D. Nelson and Moshe Givoni and Sebastian
		  Bussman},
  doi		= {10.1016/j.rtbm.2018.06.003},
  issn		= {22105395},
  journal	= {Research in Transportation Business \& Management},
  pages		= {84-97},
  title		= {User Perspectives on Emerging Mobility Services: Ex Post
		  Analysis of {Kutsuplus} Pilot},
  volume	= {27},
  year		= {2018}
}

@Article{	  Jokinen2019,
  author	= {Jani-Pekka Jokinen and Teemu Sihvola and Milos N.
		  Mladenovic},
  doi		= {10.1016/j.tranpol.2017.12.004},
  issn		= {0967070X},
  journal	= {Transport Policy},
  pages		= {123-133},
  title		= {Policy Lessons from the Flexible Transport Service Pilot
		  {Kutsuplus} in the {Helsinki} Capital Region},
  volume	= {76},
  year		= {2019}
}

@InProceedings{	  Chichung2008,
  author	= {Chichung Tao and Chungjung Wu},
  doi		= {10.1109/SOLI.2008.4682777},
  isbn		= {978-1-4244-2012-4},
  booktitle	= {International Conference on Service Operations
		  and Logistics, and Informatics (SOLI)},
  pages		= {1576-1581},
  publisher	= {IEEE},
  title		= {Behavioral Responses to Dynamic Ridesharing Services - The
		  Case of Taxi-Sharing Project in {Taipei}},
  year		= {2008}
}

@Article{	  Gargiulo2015,
  author	= {Eleonora Gargiulo and Roberta Giannantonio and Elena
		  Guercio and Claudio Borean and Giovanni Zenezini},
  doi		= {10.1016/j.promfg.2015.07.329},
  issn		= {23519789},
  journal	= {Procedia Manufacturing},
  pages		= {777-784},
  publisher	= {Elsevier},
  title		= {Dynamic Ride Sharing Service: Are Users Ready to Adopt
		  it?},
  volume	= {3},
  year		= {2015}
}

@Article{	  Zhu2021,
  author	= {Dianzhuo Zhu},
  doi		= {10.4000/rei.9984},
  issn		= {0154-3229},
  issue		= {173},
  journal	= {Revue d'économie industrielle},
  pages		= {161-202},
  title		= {The Limits of Money in Daily Ridesharing: Evidence from a
		  Field Experiment in Rural {France}},
  year		= {2021}
}

@Article{	  Kuehnel2023,
  author	= {Nico Kuehnel and Hannes Rewald and Steffen Axer and Felix
		  Zwick and Rolf Findeisen},
  doi		= {10.1177/03611981231170624},
  issn		= {0361-1981},
  journal	= {Transportation Research Record: Journal of the
		  Transportation Research Board},
  pages		= {036119812311706},
  title		= {Flow-Inflated Selective Sampling for Efficient Agent-Based
		  Dynamic Ride-Pooling Simulations},
  year		= {2023}
}

@Article{	  Fagnant2014,
  author	= {Daniel J. Fagnant and Kara M. Kockelman},
  doi		= {10.1016/j.trc.2013.12.001},
  issn		= {0968090X},
  journal	= {Transportation Research Part C: Emerging Technologies},
  pages		= {1-13},
  publisher	= {Elsevier},
  title		= {The Travel and Environmental Implications of Shared
		  Autonomous Vehicles, Using Agent-Based Model Scenarios},
  volume	= {40},
  year		= {2014}
}

@Article{	  Wilkes2021,
  author	= {Gabriel Wilkes and Roman Engelhardt and Lars Briem and
		  Florian Dandl and Peter Vortisch and Klaus Bogenberger and
		  Martin Kagerbauer},
  doi		= {10.1177/0361198121997140},
  issn		= {0361-1981},
  issue		= {8},
  journal	= {Transportation Research Record: Journal of the
		  Transportation Research Board},
  pages		= {226-239},
  publisher	= {Transportation Research Board},
  title		= {Self-Regulating Demand and Supply Equilibrium in Joint
		  Simulation of Travel Demand and a Ride-Pooling Service},
  volume	= {2675},
  year		= {2021}
}

@Article{	  Zwick2022,
  author	= {Felix Zwick and Gabriel Wilkes and Roman Engelhardt and
		  Steffen Axer and Florian Dandl and Hannes Rewald and Nadine
		  Kostorz and Eva Fraedrich and Martin Kagerbauer and Kay W.
		  Axhausen},
  doi		= {10.1016/j.procs.2022.03.079},
  issn		= {18770509},
  journal	= {Procedia Computer Science},
  pages		= {608-613},
  title		= {Mode choice and ride-pooling simulation: A comparison of
		  {mobiTopp}, {Fleetpy}, and {MATSim}},
  volume	= {201},
  year		= {2022}
}

@misc{MATSimOpenLA2020,
    title = {The {MATSim Open Los Angeles Scenario}},
    author = {{MATSim}},
    year = {2020},
    url = {https://github.com/matsim-scenarios/matsim-los-angeles}
}

@article{bilali_analytical_2020,
	title = {Analytical and Agent-Based Model to Evaluate Ride-Pooling Impact Factors},
	volume = {2674},
	issn = {0361-1981, 2169-4052},
	url = {https://journals.sagepub.com/doi/10.1177/0361198120917666},
	doi = {10.1177/0361198120917666},
	pages = {1--12},
	number = {6},
	journal = {Transportation Research Record: Journal of the Transportation Research Board},
	author = {Bilali, Aledia and Engelhardt, Roman and Dandl, Florian and Fastenrath, Ulrich and Bogenberger, Klaus},
	urldate = {2026-01-26},
	year = {2020},
	langid = {english},
}

@inproceedings{cao_sharek_2015,
	location = {Pittsburgh, {PA}, {USA}},
	title = {{SHAREK}: A Scalable Dynamic Ride Sharing System},
	isbn = {978-1-4799-9972-9},
	url = {https://ieeexplore.ieee.org/document/7264299},
	doi = {10.1109/MDM.2015.12},
	shorttitle = {{SHAREK}},
	eventtitle = {2015 16th {IEEE} International Conference on Mobile Data Management ({MDM})},
	pages = {4--13},
	booktitle = {International Conference on Mobile Data Management ({MDM})},
	publisher = {{IEEE}},
	author = {Cao, Bin and Alarabi, Louai and Mokbel, Mohamed F. and Basalamah, Anas},
	urldate = {2026-01-26},
	year = {2015},
	langid = {english},
}

@inproceedings{chen_data_2017,
	location = {Redondo Beach {CA} {USA}},
	title = {Data Driven Analysis of the Potentials of Dynamic Ride Pooling},
	isbn = {978-1-4503-5491-2},
	url = {https://dl.acm.org/doi/10.1145/3151547.3151549},
	doi = {10.1145/3151547.3151549},
	eventtitle = {{SIGSPATIAL}'17: 25th {ACM} {SIGSPATIAL} International Conference on Advances in Geographic Information Systems},
	pages = {7--12},
	booktitle = {{SIGSPATIAL} Workshop on Computational Transportation Science},
	publisher = {{ACM}},
	author = {Chen, Min Hao and Jauhri, Abhinav and Shen, John Paul},
	urldate = {2026-01-26},
	year = {2017},
	langid = {english},
}

@inproceedings{engelhardt_quantifying_2019,
	location = {Auckland, New Zealand},
	title = {Quantifying the Benefits of Autonomous On-Demand Ride-Pooling: A Simulation Study for {Munich}, {Germany}},
	rights = {https://ieeexplore.ieee.org/Xplorehelp/downloads/license-information/{IEEE}.html},
	isbn = {978-1-5386-7024-8},
	url = {https://ieeexplore.ieee.org/document/8916955/},
	doi = {10.1109/ITSC.2019.8916955},
	shorttitle = {Quantifying the Benefits of Autonomous On-Demand Ride-Pooling},
	eventtitle = {2019 {IEEE} Intelligent Transportation Systems Conference - {ITSC}},
	pages = {2992--2997},
	booktitle = {Intelligent Transportation Systems Conference ({ITSC})},
	publisher = {{IEEE}},
	author = {Engelhardt, Roman and Dandl, Florian and Bilali, Aledia and Bogenberger, Klaus},
	urldate = {2026-01-26},
	year = {2019},
	langid = {english},
}

@article{lorente_intermodal_2022,
	title = {An Intermodal Dispatcher for the Assignment of Public Transport and Ride Pooling Services},
	volume = {62},
	issn = {23521465},
	url = {https://linkinghub.elsevier.com/retrieve/pii/S2352146522001831},
	doi = {10.1016/j.trpro.2022.02.056},
	pages = {450--458},
	journal = {Transportation Research Procedia},
	author = {Lorente, Ester and Barceló, Jaume and Codina, Esteve and Noekel, Klaus},
	urldate = {2026-01-26},
	year = {2022},
	langid = {english},
}

@article{paparella_time-invariant_2025,
	title = {A Time-Invariant Network Flow Model for Ride-Pooling in Mobility-on-Demand Systems},
	volume = {12},
	rights = {https://ieeexplore.ieee.org/Xplorehelp/downloads/license-information/{IEEE}.html},
	issn = {2325-5870, 2372-2533},
	url = {https://ieeexplore.ieee.org/document/10605118/},
	doi = {10.1109/TCNS.2024.3431411},
	pages = {906--917},
	number = {1},
	journal = {{IEEE} Transactions on Control of Network Systems},
	author = {Paparella, Fabio and Pedroso, Leonardo and Hofman, Theo and Salazar, Mauro},
	urldate = {2026-01-26},
	year = {2025},
	langid = {english},
}

@inproceedings{shah_neural_2020,
	title = {Neural Approximate Dynamic Programming for On-Demand Ride-Pooling},
	volume = {34},
	rights = {https://www.aaai.org},
	issn = {2374-3468, 2159-5399},
	url = {https://ojs.aaai.org/index.php/AAAI/article/view/5388},
	doi = {10.1609/aaai.v34i01.5388},
	pages = {507--515},
	number = {1},
	booktitle = {Conference on Artificial Intelligence},
    publisher = {{AAAI}},
	author = {Shah, Sanket and Lowalekar, Meghna and Varakantham, Pradeep},
	urldate = {2026-01-26},
	year = {2020},
	langid = {english},
}

@article{shulika_spatiotemporal_2024,
	title = {Spatiotemporal variability of ride-pooling potential – Half a year New York City experiment},
	volume = {114},
	issn = {09666923},
	url = {https://linkinghub.elsevier.com/retrieve/pii/S0966692323002399},
	doi = {10.1016/j.jtrangeo.2023.103767},
	pages = {103767},
	journal = {Journal of Transport Geography},
	author = {Shulika, Olha and Bujak, Michal and Ghasemi, Farnoud and Kucharski, Rafal},
	urldate = {2026-01-26},
	year = {2024},
	langid = {english},
}

@article{soza-parra_shareability_2024,
	title = {The shareability potential of ride-pooling under alternative spatial demand patterns},
	volume = {20},
	issn = {2324-9935, 2324-9943},
	url = {https://www.tandfonline.com/doi/full/10.1080/23249935.2022.2140022},
	doi = {10.1080/23249935.2022.2140022},
	pages = {2140022},
	number = {2},
	journal = {Transportmetrica A: Transport Science},
	author = {Soza-Parra, Jaime and Kucharski, Rafał and Cats, Oded},
	urldate = {2026-01-26},
	year = {2024},
	langid = {english},
}

@article{yu_integrated_2020,
	title = {An Integrated Decomposition and Approximate Dynamic Programming Approach for On-Demand Ride Pooling},
	volume = {21},
	rights = {https://ieeexplore.ieee.org/Xplorehelp/downloads/license-information/{IEEE}.html},
	issn = {1524-9050, 1558-0016},
	url = {https://ieeexplore.ieee.org/document/8805157/},
	doi = {10.1109/TITS.2019.2934423},
	pages = {3811--3820},
	number = {9},
	journal = {{IEEE} Transactions on Intelligent Transportation Systems},
	author = {Yu, Xian and Shen, Siqian},
	urldate = {2026-01-26},
	year = {2020},
	langid = {english},
}

@article{zhu_potential_2022,
	title = {The potential of ride-pooling in {VKT} reduction and its environmental implications},
	volume = {103},
	issn = {13619209},
	url = {https://linkinghub.elsevier.com/retrieve/pii/S1361920921004508},
	doi = {10.1016/j.trd.2021.103155},
	pages = {103155},
	journal = {Transportation Research Part D: Transport and Environment},
	author = {Zhu, Pengyu and Mo, Haoyu},
	urldate = {2026-01-26},
	year = {2022},
	langid = {english},
}

@article{zwick_analysis_2020,
	title = {Analysis of ridepooling strategies with {MATSim}},
	rights = {http://rightsstatements.org/page/{InC}-{NC}/1.0/, info:eu-repo/semantics/{openAccess}},
	url = {http://hdl.handle.net/20.500.11850/420103},
	doi = {10.3929/ETHZ-B-000420103},
	pages = {15 p.},
	publisher = {{ETH} Zurich},
	author = {Zwick, Felix and Axhausen, Kay W.},
	urldate = {2026-01-26},
	year = {2020},
	langid = {english}
}

@article{zwick_ride-pooling_2021,
	title = {Ride-Pooling Efficiency in Large, Medium-Sized and Small Towns -Simulation Assessment in the {Munich} Metropolitan Region},
	volume = {184},
	issn = {18770509},
	url = {https://linkinghub.elsevier.com/retrieve/pii/S1877050921007195},
	doi = {10.1016/j.procs.2021.03.083},
	pages = {662--667},
	journal = {Procedia Computer Science},
	author = {Zwick, Felix and Kuehnel, Nico and Moeckel, Rolf and Axhausen, Kay W.},
	urldate = {2026-01-26},
	year = {2021},
	langid = {english},
}

@article{zwick_ride-pooling_2022,
	title = {Ride-pooling demand prediction: A spatiotemporal assessment in {Germany}},
	volume = {100},
	issn = {09666923},
	url = {https://linkinghub.elsevier.com/retrieve/pii/S0966692322000308},
	doi = {10.1016/j.jtrangeo.2022.103307},
	shorttitle = {Ride-pooling demand prediction},
	pages = {103307},
	journal = {Journal of Transport Geography},
	author = {Zwick, Felix and Axhausen, Kay W.},
	urldate = {2026-01-26},
	year = {2022},
	langid = {english},
}

@article{narayan_scalability_2022,
	title = {On the scalability of private and pooled on-demand services for urban mobility in Amsterdam},
	volume = {45},
	issn = {0308-1060, 1029-0354},
	url = {https://www.tandfonline.com/doi/full/10.1080/03081060.2021.2017214},
	doi = {10.1080/03081060.2021.2017214},
	pages = {2--18},
	number = {1},
	journal = {Transportation Planning and Technology},
	author = {Narayan, Jishnu and Cats, Oded and Van Oort, Niels and Hoogendoorn, Serge Paul},
	urldate = {2026-01-26},
	year = {2022},
	langid = {english},
}

@inproceedings{ota_scalable_2015,
	location = {Santa Clara, {CA}, {USA}},
	title = {A scalable approach for data-driven taxi ride-sharing simulation},
	isbn = {978-1-4799-9926-2},
	url = {http://ieeexplore.ieee.org/document/7363837/},
	doi = {10.1109/BigData.2015.7363837},
	eventtitle = {2015 {IEEE} International Conference on Big Data (Big Data)},
	pages = {888--897},
	booktitle = {International Conference on Big Data (Big Data)},
	publisher = {{IEEE}},
	author = {Ota, Masayo and Vo, Huy and Silva, Claudio and Freire, Juliana},
	urldate = {2026-01-26},
	year = {2015},
	langid = {english},
}

@inproceedings{shuo_ma_t-share_2013,
	location = {Brisbane, {QLD}},
	title = {{T-share}: A large-scale dynamic taxi ridesharing service},
	isbn = {978-1-4673-4910-9 978-1-4673-4909-3 978-1-4673-4908-6},
	url = {http://ieeexplore.ieee.org/document/6544843/},
	doi = {10.1109/ICDE.2013.6544843},
	shorttitle = {T-share},
	eventtitle = {2013 29th {IEEE} International Conference on Data Engineering ({ICDE} 2013)},
	pages = {410--421},
	booktitle = {International Conference on Data Engineering ({ICDE})},
	publisher = {{IEEE}},
	author = {Ma, Shuo and Zheng, Yu and Wolfson, Ouri},
	urldate = {2026-01-26},
	year = {2013},
	langid = {english},
}

@article{ma_real-time_2015,
	title = {Real-Time City-Scale Taxi Ridesharing},
	volume = {27},
	rights = {https://ieeexplore.ieee.org/Xplorehelp/downloads/license-information/{IEEE}.html},
	issn = {1041-4347},
	url = {http://ieeexplore.ieee.org/document/6847170/},
	doi = {10.1109/TKDE.2014.2334313},
	pages = {1782--1795},
	number = {7},
	journal = {{IEEE} Transactions on Knowledge and Data Engineering},
	author = {Ma, Shuo and Zheng, Yu and Wolfson, Ouri},
	urldate = {2026-01-26},
	year = {2015},
	langid = {english},
}

@article{danassis_putting_2022,
	title = {Putting ridesharing to the test: efficient and scalable solutions and the power of dynamic vehicle relocation},
	volume = {55},
	issn = {0269-2821, 1573-7462},
	url = {https://link.springer.com/10.1007/s10462-022-10145-0},
	doi = {10.1007/s10462-022-10145-0},
	shorttitle = {Putting ridesharing to the test},
	pages = {5781--5844},
	number = {7},
	journal = {Artificial Intelligence Review},
	author = {Danassis, Panayiotis and Sakota, Marija and Filos-Ratsikas, Aris and Faltings, Boi},
	urldate = {2026-01-26},
	year = {2022},
	langid = {english},
}

@article{liu_bus_2019,
	title = {Bus Pooling: A Large-Scale Bus Ridesharing Service},
	volume = {7},
	rights = {https://ieeexplore.ieee.org/Xplorehelp/downloads/license-information/{OAPA}.html},
	issn = {2169-3536},
	url = {https://ieeexplore.ieee.org/document/8730311/},
	doi = {10.1109/ACCESS.2019.2920756},
	shorttitle = {Bus Pooling},
	pages = {74248--74262},
	journal = {{IEEE} Access},
	author = {Liu, Kaijun and Zhang, Jingwei and Yang, Qing},
	urldate = {2026-01-26},
	year = {2019},
	langid = {english},
}

@article{dehghan_enhanced_2025,
	title = {An enhanced approximate dynamic programming approach to on-demand ride-Pooling},
	volume = {13},
	issn = {2168-0566, 2168-0582},
	url = {https://www.tandfonline.com/doi/full/10.1080/21680566.2025.2551913},
	doi = {10.1080/21680566.2025.2551913},
	pages = {2551913},
	number = {1},
	journal = {Transportmetrica B: Transport Dynamics},
	author = {Dehghan, Arash and Cevik, Mucahit and Bodur, Merve},
	urldate = {2026-01-26},
	year = {2025},
	langid = {english},
}

\end{document}